\documentclass[sigconf,screen,authorversion,nonacm]{acmart}

\AtBeginDocument{%
 }

\copyrightyear{2023}
\acmYear{2023}
\setcopyright{acmlicensed}\acmConference[AIES '23]{AAAI/ACM Conference on AI, Ethics, and Society}{August 8--10, 2023}{Montréal, QC, Canada}
\acmBooktitle{AAAI/ACM Conference on AI, Ethics, and Society (AIES '23), August 8--10, 2023, Montréal, QC, Canada}
\acmPrice{15.00}
\acmDOI{10.1145/3600211.3604682}
\acmISBN{979-8-4007-0231-0/23/08}

\usepackage{multirow}
\usepackage{hyperref}
\usepackage[nameinlink]{cleveref}

\begin{document}

\title{Bound by the Bounty: Collaboratively Shaping Evaluation Processes for Queer AI Harms}

\author{Organizers Of QueerInAI}
\affiliation{%
 \institution{Queer in AI}
 \country{}
}

\author{Nathan Dennler}
\affiliation{%
 \institution{University of Southern California}
 \country{}}
\email{dennler@usc.edu}

\author{Anaelia Ovalle}
\affiliation{%
 \institution{University of California, Los Angeles}
 \country{}}
\email{anaelia@cs.ucla.edu}

\author{Ashwin Singh}
\affiliation{%
 \institution{Queer in AI}
 \country{}}
\email{ashwin.queerinai@ostem.org}

\author{Luca Soldaini}
\affiliation{%
 \institution{Queer in AI and Allen Institute for AI}
 \country{}}
\email{lucas@allenai.org}

\author{Arjun Subramonian}
\affiliation{%
 \institution{Queer in AI and University of California, Los Angeles}
 \country{}}
\email{arjunsub@cs.ucla.edu}

\author{Huy Tu}
\affiliation{%
 \institution{Queer in AI}
 \country{}}
\email{huyqtu7@gmail.com}

\author{William Agnew}
\affiliation{%
 \institution{Queer in AI and University of Washington}
 \country{}}
\email{wagnew3@cs.washington.edu}

\author{Avijit Ghosh}
\affiliation{%
 \institution{Queer in AI and Northeastern University}
 \country{}}
\email{ghosh.a@northeastern.edu}

\author{Kyra Yee}
\affiliation{%
 \institution{Queer in AI}
 \country{}}
\email{kyrayee9@gmail.com}

\author{Irene Font Peradejordi}
\affiliation{%
 \institution{Queer in AI}
 \country{}}
\email{if76@cornell.edu}

\author{Zeerak Talat}
\affiliation{%
 \institution{Queer in AI}
 \country{}}
\email{z@zeerak.org}

\author{Mayra Russo}
\affiliation{%
 \institution{Queer in AI}
 \country{}}
\email{mrusso@l3s.de}

\author{Jess De Jesus De Pinho Pinhal}
\affiliation{%
 \institution{Queer in AI}
 \country{}}
\email{42005857@parisnanterre.fr}

\renewcommand{\shortauthors}{QueerInAI et al.}

\newcommand{\arjun}[1]{\textcolor{magenta}{[#1]}}
\newcommand{\elia}[1]{\textcolor{orange}{\textbf{[#1]}}}

\definecolor{beige}{RGB}{196, 196, 108}
\newcommand{\nathan}[1]{\textcolor{beige}{[#1]}}

\begin{abstract}
 Bias evaluation benchmarks and dataset and model documentation have emerged as central processes for assessing the biases and harms of artificial intelligence (AI) systems. 
 However, these auditing processes have been criticized for their failure to integrate the knowledge of marginalized communities and consider the power dynamics between auditors and the communities. 
 Consequently, modes of bias evaluation have been proposed that engage impacted communities in identifying and assessing the harms of AI systems (e.g., bias bounties). 
 Even so, asking what marginalized communities \textit{want} from such auditing processes has been neglected.
 In this paper, we ask queer communities for their positions on, and desires from, auditing processes.
 To this end, we organized a participatory workshop to critique and redesign \textit{bias bounties} from queer perspectives. 
 We found that when given space, the scope of feedback from workshop participants goes far beyond what bias bounties afford, with participants questioning the ownership, incentives, and efficacy of bounties.
 We conclude by advocating for community ownership of bounties and complementing bounties with participatory processes (e.g., co-creation).
 \looseness=-1
\end{abstract}

\ccsdesc[500]{Social and professional topics~Computing / technology policy}
\ccsdesc[300]{Computing methodologies~Artificial intelligence}
\ccsdesc[500]{Social and professional topics~Sexual orientation}
\ccsdesc[500]{Social and professional topics~Gender}
\ccsdesc[500]{Human-centered computing~Human computer interaction (HCI)}  
\keywords{artificial intelligence, bias bounties, harms, LGBTQIA+, participatory methods
}

\maketitle

\section{Introduction}
\label{sec:intro}

AI systems pose significant harms to marginalized communities which 
require urgent attention \cite{DiasOliva_Fighting_2021, dodge2021documenting, dev2021harms}. 
To assess AI harms, companies have used 
bias evaluation benchmarks \cite{blodgett-etal-2021-stereotyping, Felkner2022TowardsWD, ovalle2023m}, dataset and model documentation \cite{Gebru2018DatasheetsFD, Mitchell2018ModelCF, bender-friedman-2018-data}, and other auditing processes \cite{Raji2019Actionable, Raji2022Outsider, Metcalf2020AlgorithmicIA}. 
However, these processed rarely require examining the power dynamics between auditors and the communities or integrate the knowledge held in communities \cite{birhane2022power,whittaker2021steep}.
Furthermore, auditing processes are often enacted defensively by companies in response to criticism of harms from their AI systems \cite{Kayser-Bril}.

Recently, modes of bias evaluation have been proposed that engage impacted communities in identifying and assessing the harms of AI systems.
One such example is \textit{bias bounties} \cite{bar_on_2018, twitter-bias-bounty, globus2022algorithmic, bias_buccaneers}.
By including communities that have been harmed into the process of auditing systems, developers seek feedback on the types and severity of AI harms faced by those at the margins.
However, such processes fall short of allowing a full range of community feedback and control \cite{birhane2022power}. 
That is, while they may yield 
improvements, they fall short of being truly participatory approaches and can enable ethics-washing, i.e., give the appearance of 
taking steps to address ethical issues while making limited practical progress 
\cite{sloane22participation, birhane2022power}.
For example, in bias bounties, companies allow the public, often users of their systems, to interact with the systems to 
find and submit biased, toxic, or incorrect data and system outputs. 
Companies then codify and evaluate severity of harms identified using a predefined rubric.
The public rarely has a voice in how how findings are evaluated, nor do companies provide mechanisms for interrogating the \textit{internals} or existence of their data and systems.
Moreover, bounties are seldom transparent enough for participants to trace biases to design choices or structural incentives, 
let alone the efficacy to challenge the political structures in which systems are embedded.

Through this lens, one may understand current modes of participation in auditing processes as mechanisms to deny space for alternatives, thereby serving as a justification of the systems in question. 
We consider these gaps between the feedback allowed and the control mechanisms provided by auditing processes on the one hand, and what marginalized communities want on the other. 
By doing so, we seek to shift focus from what auditing processes can do to what experiences and knowledge companies allow marginalized communities to share and what companies valorize. 
We demonstrate the salience of the aforementioned gaps by conducting a participatory workshop to co-critique and redesign bias bounties from queer perspectives. 
We performed a thematic analysis on the discussions from the workshop, finding that AI systems and bounties alike pose numerous harms to queer people (e.g., exclusionary data collection, censorship, misrepresentation). 
We categorized participants' thoughts on bias bounties and systems into four main categories: 
queer harms, control, accountability, and limitations (as outlined in 
\Cref{fig:visual_results}).
In particular, participants' critiques went far beyond
how bias bounties evaluate queer harms, questioning their ownership, incentives, and efficacy.

In this paper, we 
center queer communities as 
all of the authors have done LGBTQIA+ justice work and built rapport with queer AI researchers.
We further consider bias bounties (hosted by companies to identify issues with their systems) because some authors participated in Twitter's bias bounty in 2021 \cite{twitter-bias-bounty} and were disappointed by the failure of the rubric to capture prevalent queer harms. 
As such, we intended for our workshop participants to ideate more queer-inclusive bounty evaluation rubrics.
%
Our paper argues for meaningfully engaging with marginalized communities and redistributing power to those who participate in auditing processes.
Through deeper engagements, companies can gain complex understandings of the experiences and concerns of marginalised users.
Redistributing power to participants can afford a wider range of interventions and solutions, including redlighting the use of the AI systems in question.
We further advocate for companies to engage in reflexive practices to identify the constraints placed on users, their desires, 
and examinations of the power dynamics at play.
Finally, we offer insights into data and system harms experienced by queer people and urge for community ownership of auditing processes.\looseness=-1

In particular, unless power disparities between companies and marginalized communities are minimized (i.e., communities own bias bounties), bounties \textit{cannot} be an effective auditing process. 
Bias bounties are thus \textit{incomplete} processes and are only meaningful in conjunction with other complementary steps required towards building equitable AI, e.g, co-design, and mechanisms for refusal and redress.
Additionally, regardless of ownership of the bounties, bias bounties are only applicable to the AI systems to which communities decide it is appropriate to apply bounties.

In the rest of the paper, we discuss background and related work (\S\ref{sec:related_works}), and describe our participatory and analytical methodologies (\S\ref{sec:collaborative_methods}).
We then present our workshop findings (\S\ref{sec:findings}) and discuss their implications (\S\ref{sec:discussion}).
Finally, 
we conclude our work, identify shortcomings, and provide directions for future work (\S\ref{sec:conclusion}).

\noindent \paragraph{Positionality Statement} All the authors are part of the LGBTQIA+ community. 
We are dedicated to understanding and addressing queer AI harms. 
We recognize that queer people, particularly those who are intersectionally marginalized, face unique and complex inequalities that are often overlooked in mainstream discussions of auditing. 
We further acknowledge that our positions as queer researchers in AI shape our perspectives, and we strive to be transparent about these influences in our work. 
Half the authors grew up outside the U.S. 
All authors of this paper are formally trained primarily as computer scientists. 
In addition, all authors have experience with activism, advocacy, and social work concerning queer issues. 
All workshop organizers and participants benefit from privileges which enabled them to attend our workshop. 
By collaboratively shaping auditing processes for queer AI harms, we hope to create a more inclusive and just approach to auditing that centers the experiences and needs of queer communities.

\section{Related Works} 
\label{sec:related_works}

\subsection{Queer AI Harms}
Queer people face many data and AI harms \cite{fairnessqueercomm, dev2021harms, agnew2021rebuilding, qai2023}.
Although virtual spaces are critical for queer people to find community, queer people and content are subject to increased censorship, reduced visibility, and demonetization~\cite{salty, monea2022digital, dodge2021documenting, DiasOliva_Fighting_2021}. 
Queer people also face hypervisibility, privacy violations, and surveillance, e.g., through outing via location data~\cite{time_priest, wapo_data}, monitoring on dating apps~\cite{independant_grindr}, physiognomic and essentialist attacks via machine learning~\cite{newclothes}, and invasions of online queer spaces. 
Because machine learning is preoccupied with classifying complex concepts into narrow categories, it is in tension with queerness, which can operate with concepts of fluidity of identities and seek to challenge stereotypical associations~\cite{keyes2019counting, hamidi2020gender, lu2022subverting}. 
The varied explicit risks and harms to queer people perpetuated through data and AI methods mask implicit harms, e.g., how to develop such methods 
to dismantle the structures that oppress queer and other marginalized, communities.

\subsection{Auditing Processes}
\label{sec:ethics_processes}
Several technical frameworks exist for assessing the fairness of data and AI systems (e.g., AI Fairness 360 \cite{aif360-oct-2018} and Aequitas \cite{2018aequitas}). 
However, these frameworks are based on preconceived notions of fairness situated at the system level, and do not necessarily lend to broader discourse on system design. 
Thus, they must be complemented by auditing processes that meaningfully engage with the communities impacted by the system.

Several works have called for reimagining auditing by investigating its procedural forms \cite{keyes2022feeling}; consequently, community-involved auditing processes, e.g., bias bounties, have been proposed. 
Bias bounties often consist of a company inviting communities to find and submit biases or harms in its data and systems; the company then evaluates the findings for the severity and types of harms using a predefined rubric~\cite{bar_on_2018, kenway_francois_costanza-chock_raji_buolamwini}.
For instance, Twitter held a bias bounty to uncover and assess the severity of biases in its saliency image cropping algorithm \cite{twitter-bias-bounty}. 
Our workshop participants studied and criticized the efficacy of bias bounties for surfacing queer biases and harms. 
There is a wealth of Human-Computer Interaction literature on user-driven auditing, or ``everyday'' audits \cite{hong2021everyday, devos2022toward, li2023userdriven}; these audits, like our workshop participants, strive to shift power from companies to users and stakeholders.

\subsection{Community-Based Research}

In practice, there are many challenges to incorporating modes of community participation within top-down structures, such as hierarchical companies 
\cite{Groves2023GoingPT}. 
As such, 
popular modes of participation in AI suffer from extraction and exploitation \cite{gray2019ghost} and ``participation washing'' \cite{sloane22participation}. 
Community care and diverse forms of knowledge are therefore paramount towards building just AI.
By using community-based participatory action research, research may be able to interrogate power and privilege \cite{fine2006intimate}.

Inspired by \textit{Data Feminism} \cite{d2020data} and \textit{The People's Guide to Artificial Intelligence} \cite{peoples}, our participatory workshop operationalized a ``community-first'' space for AI auditing, where queer communities were afforded space to reimagine bias bounties. 
Our workshop was a community-driven research effort in which queer facilitators (i.e., authors of this paper) invited members of the queer AI community to draw from their lived experiences to critically examine bias bounties \cite{hacker-taylor}. 
Our workshop was premised on the idea that queer researchers involving other queer researchers, as co-creators of a critical analysis of bounties, holds potential for dismantling power relations and empowering queer communities~\cite{birhane2022power, suresh22pml,kormilitzin2023participatory}. 
The resulting knowledge produced is ``by the people, for the people'' and aids in educating and mobilizing for action \cite{cooke2001participation, green2003appendix}.

\section{Methods}
\label{sec:collaborative_methods}

\begin{table*}[!t]
\centering
\caption{Participation tracks at our workshop.}
\begin{tabular}{p{0.45\textwidth} @{\hspace{1cm}}p{0.45\textwidth}} 
\toprule
\multicolumn{1}{c}{\textbf{Top-down}}                                                & \multicolumn{1}{c}{\textbf{Bottom-up}}                                                    \\ 
\midrule
\textbf{Framing:~}You are revising a framework/taxonomy to evaluate bias bounty submissions for the severity of harms discovered. & \textbf{Framing:~}You are creating a framework/taxonomy from the ground up to evaluate bias bounty submissions for the severity of harms discovered. \\\\

\textbf{Objectives:} & \textbf{Objectives:} \\

 1) Select an existing framework or taxonomy of AI harms (can be from a paper, previous bias bounty, etc.)     & 1) Select a specific AI system, and enumerate queer harms that could be introduced by this system.\\
 2) Expand upon the framework to fill gaps that pertain to intersectionally marginalized queer identities.    & 2) Find themes in these harms and develop these themes into a way of identifying, classifying, and measuring queer harms.\\
   & 3) Radically reimagine current understandings of harms and even re-envision the format of bias bounties.~   \\
\bottomrule
\end{tabular}
\label{tbl:tracks}
\end{table*}

\subsection{Participation Overview}

We held our workshop as a CRAFT session during ACM FAccT\footnote{\url{https://facctconference.org/}} (2022). 
All participants were registered as attendees of ACM FAccT (2022). 
We invited participants to form teams to develop holistic and inclusive evaluation guidelines for queer AI bias identification, measurement, and categorization and propose best practices for auditing AI systems for queer biases. 
All participants volunteered for the workshop and were made aware of it through the FAccT program and posts on Twitter.
Participants were given two key research questions to consider:\footnote{All details provided to the participants are provided in the supplementary material.}
\begin{enumerate}
  \item Where can frameworks for understanding AI harms be expanded to encompass queer identities? 
  \item How can the lived experiences of queer people inform the design of harm evaluation frameworks? 
\end{enumerate}

Participants were encouraged to consider a 
variety of AI systems, e.g., text, speech, images, graphs, tabular data, and how these systems interact with and affect queer people. 
We hosted two separate three-hour sessions: a virtual session and an in-person session.

\subsubsection{Team Formation} 
Participants self-organized into teams in each session. 
Contributors had the chance to opt into a matching program to be paired with other workshop participants. 
We requested teams to be interdisciplinary, for which reason participants sought members with different research backgrounds. 
Across the two sessions, there were nine teams with approximately 3-5 participants per team; six teams participated in the in-person workshop and three teams participated in the virtual workshop.
Each team was joined by a facilitator (i.e., an organizer of the CRAFT session), who supported and guided the team. 
Each team also designated a recordkeeper of its discussions.
All participants were invited to share their process, experiences, and thoughts. 
In our thematic analysis of participants' discussions (\S\ref{sec:findings}), we only include the work of participants who provided affirmative consent for us to do so. 

\subsubsection{Approaching the Critique} 
Teams were provided with two approaches to reimaginging bias bounties: a top-down or bottom-up approach. 
In the top-down approach, teams were encouraged to critique how bounties currently evaluate harms while in the bottom-up approach, teams considered harms that AI systems pose to queer people and used these harms as a grounding to re-envision bounty design. 
Each track came with examples, literature, and guiding questions to help teams get started (c.f., supplementary material). 
For instance, 
for the bottom-up track, we provided various example AI systems to be critiqued, such as the AllenNLP demos \cite{Gardner2017AllenNLP}, AI dungeon \cite{ai_dungeon_2019}, OpenAI's DALL-E \cite{ramesh2021zero}, and GLIDE \cite{nichol2021glide}. 
For the top-down track, we provided examples of queer AI harm ontologies, such as \citet{salty} and \citet{dev2021harms} . 
\Cref{tbl:tracks} summarizes the top-down and bottom-up tracks and their objectives.

\subsubsection{Consent and Rapport-Building}
We did not seek IRB approval for our workshop due to the difficulty of 
approvals 
recognized across every participating geography, university and company that the authors represent. 
However, the proposal for our workshop was reviewed and approved by the FAccT CRAFT chairs, and participants were informed of the format, benefits, and risks of the workshop ahead of time. 
Participants also filled out a form to express their consent to have their work included in our analysis.

We provided attendees with a code of conduct and an anti-harassment policy, emphasizing the protection of the privacy and safety of all individuals at our workshop. 
We motivated the workshop by providing background on bias bounties and the hegemonies underlying AI systems that inevitably lead to a lack of trust in them and companies. 
We further provided scholarly case studies and articles on queer AI harms (e.g., misgendering, erasure, outing). 
While we did not explicitly document how many attendees identify as LGBTQIA+, such harms reflected a shared reality of many attendees, who were open about how they identify.

\subsubsection{Participant Positionality} 
By hosting our workshop at ACM FAccT (2022), with the organizers and participants in the same community, we aimed to minimize power disparities between the organizers and participants. 
Building rapport with participants is a common practice in ethnographic research \cite{glesne1989rapport}, as it provides support for participants disclosing potentially sensitive experiences. 
This is particularly important in the context of AI harms, as many queer people have experienced social, emotional, and psychological distress \cite{fairnessqueercomm, powell2020digital, noaccess, bulliedblackmailed}. 
However, given that all our workshop participants were FAccT attendees, their views reflect a particular positionality: one that has access to resources to attend the conference, is generally associated with an institution, is English-speaking, and has the technical literacy required to scrutinize AI systems. 
Joining our workshop also indicated 
that attendees were comfortable with being in visible proximity to LGBTQIA+ spaces. 
The views of those who 
outside this positionality are less likely to be reflected in our analysis.

\subsection{Thematic Procedures}
All teams converged to similar critiques of and recommendations for bias bounties, regardless of the track in which they participated. 
We therefore consider 
all discussions collectively 
rather than perform 
separate analyses for each track. 
After the workshop, we conducted an iterative inductive thematic analysis of the participants' discussions, following 
\citet{clarke2015thematic}. 
We use this interpretivist approach to surface how queer populations desire bias bounties to be implemented. 
We used the following process:
(1) we compiled all 
submitted artifacts from the workshop, that participants consented to being analyzed into a single document, 
(2) each researcher independently developed codes for all artifacts in the document, 
(3) researchers collaboratively sorted these codes into initial themes, 
(4) concepts were grouped into overarching themes and sub-themes, and 
(5) steps 3-4 were repeated with different subsets of researchers until all researchers agreed on a 
set of themes. 

\section{Thematic Analysis Findings}
\label{sec:findings}

\begin{figure*}
  \centering \includegraphics[width=\linewidth]{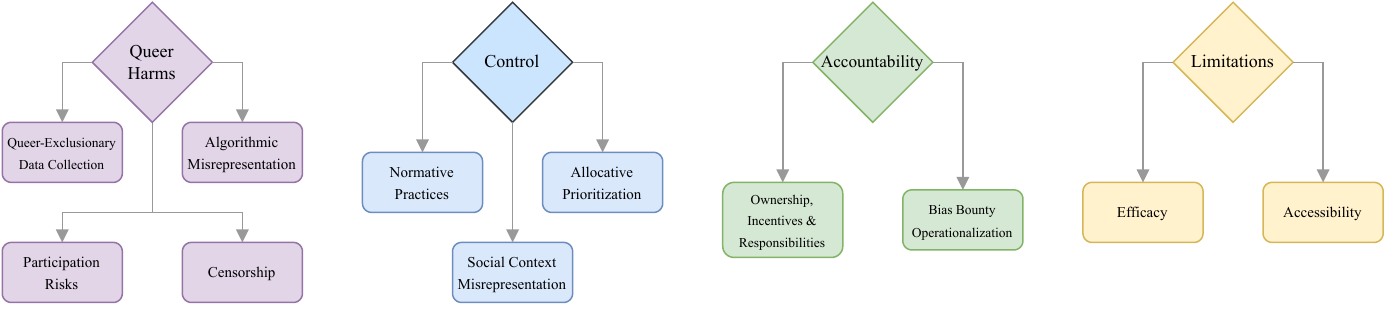}
  \caption{Taxonomy of participant critiques of AI systems and bias bounties.}
  \label{fig:visual_results}
\end{figure*}

In this section, we present the findings of our thematic analysis.
We found that our workshop participants discussed how bias bounties pose harms to queer people, in addition to the harms posed by AI systems. 
The participants' thoughts on bias bounties and systems fell into four main categories: queer harms, control, accountability, and limitations, as outlined in Figure \ref{fig:visual_results}. 
We provide a table of team frequencies for each sub-theme in \Cref{tab:counts}. 
For clarity, we use \textbf{P} to denote teams that participated in person; all in-person teams selected the bottom-up track. 
We denote teams with a \textbf{V} if they participated virtually; all virtual teams selected the top-down track.

\begin{table}[]
\caption{Number of teams that discussed each sub-theme.}\label{tab:counts}
\begin{tabular}{llc}
\hline
Theme & Sub-theme & Teams\\ 
\hline

\multirow{4}{*}{\textbf{Queer Harms}}                               & \begin{tabular}[c]{@{}l@{}}Queer-Exclusionary \\ Data Collection\end{tabular}      & \textbf{5} \\
                                             & Algorithmic Misrepresentation                              & \textbf{7} \\
                                             & Participation Risks                                   & \textbf{6} \\
                                             & Censorship                                        & \textbf{4} \\ \hline
\multirow{3}{*}{\textbf{Control}}                                 & Normative Practices                                   & \textbf{4} \\
                                             & Allocative Prioritization                                & \textbf{2} \\
                                             & \begin{tabular}[c]{@{}l@{}}Social Context \\ Misrepresentation\end{tabular}       & \textbf{3} \\ \hline
\multirow{2}{*}{\textbf{Accountability}}                             & \begin{tabular}[c]{@{}l@{}}Ownership, Incentives, \\ and Responsibilities\end{tabular}  & \textbf{2} \\
                                             & \begin{tabular}[c]{@{}l@{}}Bias Bounty Operationalization \\ Considerations\end{tabular} & \textbf{2} \\ \hline
\multirow{2}{*}{\begin{tabular}[c]{@{}l@{}}\textbf{Limitations of} \\ \textbf{Bias Bounties}\end{tabular}} & Efficacy                                         & \textbf{3} \\
                                             & Accessibility & \textbf{3}\\
                                             \hline
\end{tabular}

\end{table}

\subsection{Queer Harms}
All teams ($n=9$) identified several harms that affect how queer people are represented in and interact with AI systems and bias bounties. 
We refer to these harms as ``queer harms,'' because they are directly tied to users' queer identity.

\subsubsection{Queer-Exclusionary Data Collection}
\label{sec:queer_exclusionary_data_collection}

Several teams ($n=5$) were concerned about how queer people are represented in data in the context of both AI systems and bias bounties. 
Regarding AI systems, participants discussed how queer intersectional identities may constitute a smaller part of a user base, which can lead to harm from systems that are trained on user data. 
\textbf{P4} summarized this as ``intersectional subgroups are not well represented in the data'' and 
elaborated that this can also lead to, e.g., failures of content filters that aim to reduce exposure to harmful content because ``slurs used by small groups, e.g. specific hate groups, are not detected,'' exposing vulnerable users to psychologically harmful content.

Teams additionally discussed ways that AI systems can harm queer users by collecting personal data, especially user gender. 
Participants found that collecting gender data requires highly contextual considerations. 
\textbf{P1} noted that ``members of the queer community find it empowering and affirming to link their identity to existing socially salient categories,'' but also warned that in a clothing recommendation context, asking users to provide gender information could ``reinforce the possibly harmful idea that clothes are gendered.'' 
\textbf{P6} argued that users self-reporting gender is better than companies inferring gender, which reduces user autonomy. 
\textbf{P1} posed the question, ``How much control do you have in terms of what category you're applied to (e.g., Computer Vision fields based on external 'passing' elements vs. What content do you consume/produce?),'' indicating that the 
autonomy to fully describe 
ones own identity is a key concern for queer users.

While bias bounties were created to allow the communities impacted by particular AI systems to address such tensions, participants pointed out that bias bounties suffer from similar issues as AI systems, i.e. queer participants in bias bounties still feel marginalized. 
As \textbf{P1} noted, ``queer users are not a majority, this may already be the root cause of some issues.'' Participants questioned how effective bias bounties could be at addressing the concerns of populations that are not represented by the majority. 
\textbf{P6} asked, ``Whose interests are we optimizing on? [...] The majority? Specific groups?'' 
If bounties use sample size to determine the severity or prevalence of harms, rather than social and historical context, biases that affect queer people may receive less focus than biases affecting larger groups. 
Overall, participants were concerned that bias bounties may not operate with their interests in mind, especially when their processes are not transparent.

\subsubsection{Algorithmic Misrepresentation} 
\label{sec:algo-misrepresent}
Teams ($n=7$) voiced a concern over how data representation reduces the complexity of identity, thereby enabling their erasure and oppression.
Participants identified two ways in which this can happen: (1) when categorical representations do not capture their identity at all, and (2) when representations do not allow changes over time.

Participants expressed frustration at how categorization constrains how queer users express their identity, in particular, highlighting the tension between how ``users themselves identify \& record their identity'' (\textbf{P6}) and how identity may be represented in systems. 
Teams noted that systems that to handle diverse users and try to operationalize identity as part of their user experience can inhibit marginalized users from expressing their identity:

\begin{quote}
  \textbf{\textbf{P2}:} ``Highly structured and normative processes, typically don't have space for queer and intersectional identities which can make the process challenging if not inaccessible.''
\end{quote}

Other participants voiced that categorization ``reinforces the existence of gender categories more broadly (which some members of the queer community find inherently oppressive)'' (\textbf{P1}), which creates a ``boundary box of having/not having an identity being aggressively reinforced'' (\textbf{P6}), and ``could lead to erasure [of marginalized identities]'' (\textbf{P5}). 
For instance, companies may force individuals to assign themselves to categories, or even infer categories, for reasons including ``aggressive [content] recommendation and promoting specific content'' (\textbf{P6}). 
Beyond ``being profiled/categorized automatically into something you are not'' (\textbf{P1}), forced categorization can stereotype users, e.g., via ``recommendation of jobs that reinforce certain assumptions about your identity.'' (\textbf{P5}).
Forced categorization can further exnominate queerness:
\begin{quote}
  \textbf{\textbf{P6}:} ``People might have wrong perceptions of what it means to have a specific identity.''
\end{quote}

The appropriateness of categorization is highly contextual.
For example, not considering gender labels in a clothing recommendation system could be one solution to address the harms of categorization. 
Another solution can be having a customizable gender input, which may allow users to feel affirmed in their identity, as noted by \textbf{P1}. 
Furthermore, categorization may be required for personalized content moderation (\textbf{P6}).

In addition to being able to accurately have one's identity represented at a given point in time, teams ($n=4$) expressed the desire to be able to \textit{change} how one's identity is represented over time, \textit{as} their identity changes.
For example, \textbf{P5} noted that ``categorization does not recognize fluidity of labels over time,'' and \textbf{P6} echoed that static categories lose relevance over time because ``using old data won't represent you accurately.'' However, \textbf{P2} commented that friction in changing personal information can have organizational costs that can potentially frustrate or deter users from platforms: 

\begin{quote}
  \textbf{\textbf{P2}:} ``Name changes or pronoun changes make admin much harder and more expensive through time, complexity, or financial penalty.''
\end{quote}

Overall, teams called upon companies to build AI systems without the assumption that a user can be accurately represented in perpetuity as when they start using a system.

\subsubsection{Participation Risks}
\label{sec:participant-risk}
Throughout the workshop, many teams ($n=6$) discussed how 
using AI systems and participating in bias bounties may 
negatively impact queer people. 
The teams identified two primary risks: 
(1) the increased likelihood of personal information being disclosed, and (2) increased exposure to 
harms. 
\textbf{V3} described these harms succinctly as ``privacy harm'' and ``exposure harm.'' 
Privacy harms referred to AI systems and bias bounties, while exposure harms were unique to participation in bias bounties.

Regarding AI systems, participants highlighted the risks of personal information being disclosed and people being outed. 
For example, \textbf{V3} and \textbf{P1} noted that Grindr had shared HIV status with third parties, TikTok had censored and surveilled LGBTQ language, and surveillance software had reported personal Google Docs data to schools. 
\textbf{P6} commented, ``What if you don't want to be more exposed/visible? Queer people can be exposed to harassment if they get too much visibility then they might not want to be exposed to a larger audience.'' 
This highlights that undesired visibility can cause queer people to face harm, discrimination, and harassment.

Participants also discussed how their personal data can be disclosed via their participation in bias bounties. 
\textbf{V1} noted that bounties can ``put you in a dangerous situation, [such as] forced outing,'' if you submit queer harms you have experienced. 
In addition to outing, \textbf{V3} stressed that privacy concerns should motivate the organizers of bounties to ensure ``anonymous reports of non-anonymous content/interactions,'' e.g., through the removal of personally identifiable information. 
In particular, \textbf{V3} explored how specific queer populations (e.g., queer youth) may need extra safeguards in place to ensure safe participation, such as ``consent from parents.'' 
Moreover, there is potential for adversarial attacks on bounties, such as hate crimes targeting queer groups by changing their profiles or flooding systems with disturbing content. 
Overall, participants emphasized that bias bounties should provide participants with warnings and safeguards to prevent their personal data from being disclosed to wider audiences.

Teams also indicated that bias bounties pose ``exposure harm'' to their participants through the identification and submission of negative interactions with systems.
. 
This increases bounty participants' exposure to psychologically harmful content. 
\textbf{V2} asked, ``Are bug bounties exploitative? People have to experience biases,'' and \textbf{V3} similarly noted that bounties facilitate ``exposure to sensitive topics.'' 
These teams noted bounties currently subject participants for marginalized communities to witness how AI systems mistreat them and members of their community.

\subsubsection{Censorship}
\label{sec:censorship}

Our analysis revealed the experiences and desires of queer users in relation to AI harms and their resilience in navigating these harms in efforts towards reimagining systems. 
For example, participants expressed frustration with how online harassers exploit the weaknesses of content moderation AI (e.g., failure to consider context) via dog whistles, with \textbf{P4} noting that ``obfuscated hate speech'' often goes undetected or ignored. 
At the same time, queer users, with their resilience, co-opt this failure mode by self-censoring to protect themselves and their content from surveillance, fetishization and sexualization, as expressed by \textbf{P6} and \textbf{V3}. 
Participants came up with innovative ways to bypass censorship (e.g., \textbf{P6} replaced the word ``sex'' with ``seggs,'' and \textbf{V3} described that ``lesbian'' is often replaced with ``le\$bian'').

However, the lack of explainability of content moderation AI makes it challenging to contest censorship and amplifies harms. 
\textbf{P6} highlighted the absence of explanations for recommendations and bans on dating apps; but, queer users' resilience leads to the possibility of reimagining AI systems as a reality, e.g., systems can prioritize human-in-the-loop explainability (\textbf{P6}).

Participants also emphasized a desire for their identities to be recognized, rather than erased or ignored, by AI systems.
For instance 
\textbf{P1} found it ``empowering and affirming to link their identity to existing socially salient categories.''


\subsection{Control}
\label{sec:control}
Participant teams ($n=5$) raised 
concerns about who controls bias bounties and the AI systems that are the subject of bounties due to its impact on the degree of access granted to bounty participants.

\subsubsection{Normative Practices}
\label{sec:normative_practices}
Participants noted that there are norms encoded in systems that harm queer users. 
For instance, \textbf{P4} stated, ``Community standards encode perspective of the privileged groups, not the marginalized groups.'' 
Thus pointing towards a misalignment 
of values between 
companies and users, 
resulting in ignorance and exploitation by companies, intentional or otherwise. 

Several teams provided specific examples of how the values of companies and users can be misaligned. 
In the context of financial services, \textbf{P2} shared that ``payday loan companies target vulnerable populations,'' which can benefit the company by fulfilling key metrics but harms its users via predatory practices. 
In AI hiring contexts, \textbf{P5} noted that ``groups are selected at different rates,'' which can exacerbate the ``underrepresentation of marginalized groups.'' While these systems might make the 
recruitment process easier for companies, they harm the groups that are under-represented. 
With regards to social media, \textbf{P6} said, ``Optimization based on engagement can increase harassment.'' 
Engagement may increase time spent on the platform, benefiting the company, but toxic engagement can harm marginalized users. 
Without actively considering users when optimizing metrics, 
companies risk harming their users.

Participants pointed out that this critique also applies 
to companies that run bias bounties. 
Participants expressed concern about how companies may be misaligned with bounty participants. 
\textbf{P5} asked, ``Who gets to decide the `normative' notion of fairness?'' 
Companies following normative practices when running bias bounties may overlook ways of improving systems that can actually benefit users; furthermore, metrics (e.g., to measure the severity of harms) that bias bounties aim to optimize may result in unforeseen harms.
\textbf{P5} asked, ``should an algorithmic / AI system even be used?'' Bias bounties cannot accommodate such a radical systemic change, because they instead focus on reforming parts of systems. 
This raises a need to fundamentally reconsider how companies build AI systems, and if and when to do so at all. 
\textbf{P5} further contextualized their question in the domain of hiring: ``Should algorithms even be used in hiring? If no, [that] assumes that you can't improve upon the status quo.'' 
This highlights the tension between new and existing processes; even if new processes do not benefit impacted or marginalized communities, it is not guaranteed that existing processes are free of harms.

\subsubsection{Allocative Prioritization}
\label{sec:allocative_prioritization}
Teams ($n = 2$) expanded on how normative practices can result in imbalanced resource allocation, which can result in urgent harms to 
users. 
\textbf{P1} stated that ``some harms might be life-threatening.'' 
Providing an example, \textbf{P2} remarked that 
``life-saving surgeries (e.g. transition) are unsecured and/or considered elective and so harder to secure credit than [for] more traditional items e.g. car[s].'' 
AI systems that allocate resources may not prioritize queer concerns, and thus make system usage disproportionately more difficult and harmful for queer users.

Participants found that bias bounties also risk unfairly distributing resources to queer people.
\textbf{P2} commented that ``financial services organizations tend to lack diversity which can make it challenging to begin conversations to understand queer harm, nevermind evaluate it.'' 
Without queer representation in the company running a bounty, queer harms may not even be represented in the bounty rubric, let alone identified.

\subsubsection{Social Context Misrepresentation}
\label{sec:social_context}
Teams ($n=3$) emphasized that social context is critical to ensure that AI systems and bias bounties benefit users. 
Regarding AI systems, participants noted how misrepresenting users' social context can negatively impact users. 
For example, in content moderation contexts, \textbf{P4} noted that ``non-English language [are] not well handled (automatic translation errors, lack of data);'' if systems incorrectly process languages other than English, they can expose their users to toxic content. 
\textbf{P6} also noted that ``different identities might be connected to different levels of maturity/safety during content moderation.'' 
That is, when systems are unable to adapt to the social contexts of different age groups, they also harm their users.

Social context is also critical for bias bounties. 
\textbf{P6} stressed that ``accommodating cultural diversity'' is a criterion that bounties must satisfy in order to be effective. 
However, \textbf{P6} qualified this by asking, ``How much cultural difference makes sense? How much of it is represented in the technology? Is it useful to increase diversity or granularity?'' 
Among participants, there was a recurring theme of seeking 
the right balance between representing social context with enough detail to capture the diversity of the systems being audited, but not so much detail that participation in the bias bounty becomes infeasible.

\subsection{Accountability}
Some teams ($n = 3$) discussed specifics of how bias bounties should be run. 
These teams stressed the importance of community-led over company-led bounties.

\subsubsection{Ownership, Incentives, and Responsibilities}
\label{sec:ownership_incentives_responsibility}
Teams ($n=2$) expressed concern about the ownership, incentives, responsibility of companies organizing bias bounties.
For instance, \textbf{V3} stated that ``we don't want random tech companies to have ownership of this,'' and \textbf{V1} asked, ``how do we know that the distribution of data being handled to us is not adversarially generated?'' 
Participants argued 
that companies lack incentives to run bias bounties due to: (1) misalignment with companies' values (\textbf{V3}: ``it might not align with company legal framework''); (2) harmed users not being a majority (\textbf{V1}: ``queer folks are a `small group' at the margins and don't bolster overall utility or revenue maximization''); and (3) financial hurdles for companies (\textbf{V1}: ``any kind of audit costs money''). 
Worse, \textbf{V1} highlighted that companies are often disincentivized to uncover harms ``because of legal risks.'' 

To address issues of ownership, incentives, and responsibility, participants suggested: (1) employing a trusted third party and partnerships with local, trusted organizations to mitigate concerns (\textbf{V3}: ``Partnership with local youth centers as data stewards... Tech companies may provide tooling, not governance''); (2) drawing upon existing audit mechanisms from other fields, e.g., software development (\textbf{V1}: ``mechanisms for auditing exist in software development''); and (3) shifting incentives for companies to prioritize ethical considerations and participate in audits. 
Ultimately, mitigating queer AI harms requires ongoing involvement from and ownership by queer communities to ensure that ethical considerations are prioritized over companies' legal and financial risks, and align with the values and needs of the communities.

\subsubsection{Bias Bounty Operationalization}
A central question of our work is: ``How do queer communities imagine bias bounties?'' 
Participants expressed their dissatisfaction with the current format of bias bounties and envisioned a new, community-based bounty format: 
``we envisioned it as a collaborative bug bounty where individuals can contribute with specific examples towards identifying harms'' (\textbf{V2}).  
This community-based approach would involve a coalition formed by researchers and impacted communities, with communities having the power to veto AI systems entirely.

Participants stressed the importance of diversity in the operationalization of bias bounties. 
For instance, \textbf{V1} emphasized the need for diverse data collection to robustly evaluate bounty findings and better understand the distribution of system use cases, and called for harm mitigation mechanisms beyond bias bounties, e.g., community focus groups and distributed AI developments.
As an example, \textbf{V1} called for creating a focus group comprising of annotators, developers, and bounty participants from diverse backgrounds to provide feedback ahead of system development.
\textbf{V1} further posited, ``community-led AI could reduce how much context is lost via centralization and scale.''
Thus, our participants highlighted the need for community ownership of auditing processes as a means to create bounties that emphasize the needs and experiences of queer communities.

\subsection{Limitations of Bias Bounties}
Teams ($n=3$) also reflected on the appropriateness of bias bounties for addressing AI harms, discussing their effectiveness and ease of participation. 

\subsubsection{Efficacy}
\label{sec:efficacy}
Even in an ideal scenario, 
where the implementation of bias bounties poses no risks to participants, teams ($n=3$) still were wary of how helpful bounties can be.
As \textbf{V1} highlighted, ``often the answer is not here is an improvement to model, but rather you should not be doing this... this is a harder answer for people to stomach.'' 
This goes beyond identifying individual harms via bias bounties, to addressing the root causes of harms and questioning the existence of some technical systems entirely. 

Bias bounties are also challenged by the difficulty in evaluating the severity of harms. 
Such evaluations often depend on individuals and their particular subjectivites and contexts.
For instance, as one participant explained, ``Child Protection Service rips marginalized children away from families... Very harmful, but people in the system claim they are doing good. Make a framework for creating a harm severity framework instead of a single harm severity framework.'' 
Thereby suggesting that perspective and political motivations influence the lens through which a harm severity is viewed.
Moreover, the participant's example illustrates that 
current auditing processes do not adequately capture the complexity and diversity of harms that marginalized communities face, therefore necessitating a new approach to developing frameworks for harm severity. 

\textbf{V2} highlighted a tension between intent and impact, arguing that there are ``many things developers should think about, but they can't anticipate everything.''
Thus, even with good intent in mind, bias bounties fail to capture the full range of harms that marginalized communities experience, and must therefore be adaptable to harms that were not expected, or in the words of \textbf{V3}: 
``[we] don't want to lock in stone the current ways we think about LGBTQ harms and interactions, need to leave room for growth.''
 
While bias bounties have their limitations, participants acknowledged their value as a way to decentralize the process of identifying biases and harms. 
As \textbf{V1} noted, ``bias bounties are nice ways to open-source process of identifying biases and harms.'' 
At the same time, bias bounties are not a one-size-fits-all solution for mitigating harms. 
As \textbf{V1} pointed out, while a ``bias bounty does not consider ways to address biases,'' ``reactive systems are still an important part of the problem since we know that we can't catch everything from the get-go.'' 
However, participants found that having a variety of ways to provide feedback to AI systems, in addition to bias bounties, would help make systems more beneficial to users. 
For example, \textbf{V1} questioned, even with community-owned bias bounties, ``Will the same power structures be replicated? Are we just pushing the problem downstream?''; devising additional avenues for reactively mitigating AI harms can help communities better iteratively co-develop systems. 

\subsubsection{Accessibility}
\label{sec:accessibility}
Multiple teams ($n=3$) noted that the potential learning curve to participating in bias bounties poses a barrier to many communities. 
Particularly, \textbf{V1} remarked on the technical difficulty that ``[a bias bounty] requires technical know-how'' there needs to be a ``Way to make it accessible to broader community without a technical background.''.
Another participant, \textbf{V2} noted that there is a ``barrier to entry: [bounties are] weighted towards technical people.'' 
That is, not only does the required technical knowledge pose difficulties in using systems, but it may also directly exclude entire communities from participating, which skews who can provide input on how AI systems should work and thus, for whom bounties can provide change. 
Beyond providing instructions on how to use systems, bias bounties should ``educate on the risks of sharing data, show ways to minimize sharing of personal information, discuss participation with parents/guardians, [describe] general internet security etiquette, [provide a] history of how community activism has been effective in the past'' (\textbf{V3}).
Such education can be compiled into a ``digital toolkit'' that is provided to bounty participants during onboarding (\textbf{V3}); such a toolkit could minimize the risks of participating in bounties, especially for marginalized communities that face disproportionate risks of harms.
%
%

\section{Discussion}
\label{sec:discussion}
\subsection{Enabling Interventions Throughout the AI Pipeline}
Throughout the workshop, participants reimagined auditing processes to address queer AI harms. 
We found that participants desired interventions at all stages of the AI pipeline: 
system formulation, data collection for the system, system design and development, and recourse after it has been deployed. 
Currently, many processes for addressing harms (e.g., bias bounties) are only used at the final stage of the pipeline, after the system has been deployed.
We explore how the findings from our workshop apply to different stages of the AI pipeline, and discuss potential 
interventions at each stage.

\subsubsection{Problem Formulation}
Participants desired mechanisms to provide feedback on the intended application domain of a system before it is developed (\S\ref{sec:normative_practices}). 
Furthermore, participants described three areas for 
feedback from queer communities before the system is implemented: (1) assessing the applicability of normative practices in a new context (\S\ref{sec:normative_practices}), (2) determining how resources should be allocated (\S\ref{sec:allocative_prioritization}), and (3) clarifying the incentives that drive the system's development (\S\ref{sec:ownership_incentives_responsibility}). 
One solution participants suggested was to organize a community-based panel to assess AI systems; 
the panel's goal is to proactively identify potential harms by engaging with members of queer communities. 

\subsubsection{Data Collection}
Participants also desired mechanisms for providing feedback on data collection procedures, particularly to prevent: the (1) 
exclusion of intersectional queer identities during data collection (\S\ref{sec:queer_exclusionary_data_collection}), and (2) 
misrepresentation of queer communities' social contexts (\S\ref{sec:social_context}). 
Participants indicated that they 
desired transparency around the composition of company data, especially as it concerns the representation of queer people. 
Participants desired full control over how they are represented in data, towards dismantling constraints on their expression and the fluidity of their identity.
They also expressed a desire for companies to invest in understanding the contexts in which queer people are susceptible to being outed, 
misgendered, 
censored, and experiencing dysphoria (\S\ref{sec:allocative_prioritization}, \S\ref{sec:social_context}). 
It is therefore paramount that companies engage in long-term, cooperative relationships with harmed communities, and relinquish control of data auditing processes to them.

\subsubsection{Algorithm Development}
\label{sec:algo-dev}
Participants raised concerns about inaccurate representations in AI decision-making and a loss of autonomy through AI censorship (\S\ref{sec:algo-misrepresent}, \S\ref{sec:participant-risk}, \S\ref{sec:censorship}).
Largely, participants wanted ways to be precise in how they express their identity and explanations for system behavior. 
Participants wanted to leave the choice to disclose specific aspects of identity to users themselves, rather than disclosure being a requirement to use a system. 
Beyond this, participants wanted the freedom to change how they represent themselves over time. 
These concerns echo the principles of designing with affirmative consent~\cite{strengers2021can, im2021yes}; in particular, ensuring that persons disclosing their identity are informed of what disclosure means,
being able to disclose their identity precisely and freely, and being able to reverse disclosure without consequences.

\subsubsection{Deployment}
While bias bounties afford
identifying biases in deployed AI systems, participants commented on the risks and accessibility of participating in bounties (\S\ref{sec:participant-risk}, \S\ref{sec:accessibility}). 
Participants desired transparency about the ownership, incentives, and efficacy of bounties (\S\ref{sec:normative_practices}, \S\ref{sec:ownership_incentives_responsibility} \S\ref{sec:efficacy}). 
Participants also noted the vital importance of informed consent to participate in bounties.
A team suggested providing digital toolkits to bounty participants to broaden participation in AI audits while reducing the likelihood of adverse effects. these toolkits were imagined to consist of tutorials, information, and guidelines on how to safely and correctly engage in community-involved audits of AI systems. 

\subsection{Scrutinizing Bias Bounties}
We found that participants in our workshop frequently wanted greater power in shaping bias bounties (\S\ref{sec:control}), including defining how submissions were evaluated, how compensation would be disbursed, who would provide funding, and who would organize and control the bounty. 
Members of a community that frequently experience data and AI harms could quickly identify constraints on the feedback they were able to provide about systems. 
This finding reveals key weaknesses of bias bounties: what gets counted as a harm, the need for funding for prizes, and control of the bounty by companies often leading to late-stage interventions focused on improving the system rather than reimagining how the system should have been created. 

We argue that this is a useful lens for understanding the limitations of auditing processes. 
Harmed communities should not just be participants in auditing processes, but also be asked what they think about the processes in the first place, and ideally be involved as co-designers and owners of the processes. 
Auditing researchers often valorize expert knowledge and institutional audits; however, holding auditing processes to the standard of scrutiny by harmed communities will ground them in the actual needs and realities of the people who need effective auditing the most.

Moreover, we argue that unless power disparities between companies and marginalized communities are minimized (i.e., communities own bias bounties), bounties \textit{cannot} be effective. 
In addition, even if communities own bounties, communities (not companies) should decide if a bounty is even appropriate for identifying the harms of certain AI systems. 
Ultimately, bias bounties are \textit{incomplete} processes---they are merely one of numerous complementary steps that companies need to take towards building equitable AI (e.g., co-design, mechanisms for refusal and redress). 
While bias bounties can be valuable for uncovering harms in AI systems that may not have been foreseen during development, harm anticipation, identification, and mitigation must begin outside bounties. 

\subsection{Imagining Community Ownership of Auditing Processes}
\label{sec:community_ownership}
Many participants questioned the ownership of bias bounties, discussing how corporate ownership of bounties is a conflict of interest that may lead to misaligned incentives. 
When imagining solutions, participants repeatedly mentioned empowering harmed communities, often to the point of giving them control or ownership of the bounty. 
This insight addresses several issues of auditing processes, while also presenting novel sociotechnical challenges for auditing researchers. 
First, giving harmed communities ownership of auditing processes ensures that the incentives of auditors align with the values and needs of the communities the processes are intended to help. 
Second, community ownership increases trust in the auditing. 
Third, community ownership allows for auditing processes to prioritize and be adapted to small communities that may be sparsely represented in broader audits. 
Ultimately, harmed communities understand their issues best, and are thus best positioned to conduct audits of the AI systems that impact them.

We now concretize what community-owned bias bounties might look like. 
External elements, like a public competition to find instances of bias in systems for prizes, would remain. 
Companies may voluntarily provide their system and even funding for prizes. 
However, we believe that bounty organizers will often have to contend with auditing closed-source systems and obtaining their own funding. 
While API access alone can be sufficient for auditing and redteaming closed-source systems~\cite{rando2022red, wang2023robustness}, many systems lack public API access, and API owners may take countermeasures to detect and prevent adversarial use. 
Developing methods to probe systems with limited access presents interesting directions for auditing research. 
Furthermore, marginalized communities often do not have the same financial resources as companies \cite{qai2023}. 
State grants and external non-governmental organizations may provide funding; however, these entities may have misaligned incentives and goals, in addition to requiring specific networks to access their funding. 
We argue for making community-owned bias bounties financially sustainable by reimagining bounty work, and more broadly resisting and fighting data and AI harms, as a form of mutual aid~\cite{spade2020solidarity}. 
Specifically, bias bounty work should be motivated and sustained by the direct and positive impact it has on harmed communities. 
Auditing processes often require several 
years of work to coordinate different experts and institutions to achieve noticeable change; in contrast, community-owned 
auditing processes can have more immediate and direct impacts. 
Creating community-owned auditing processes would require asking new questions in auditing research, such as what expertise and resources are needed for a bounty or audit, and how they can be made accessible to harmed communities, as well as how the impact these processes have can be made visible and tangible to motivate their usage.

\section{Conclusion}
\label{sec:conclusion}

While many auditing processes exist for identifying AI biases and harms, current operationalizations thereof are hierarchical and reflect an epistemic authority; the companies that ask for critical feedback are the same companies which force marginalized communities to comply with their definitions, parameters, and guidelines around harms that may not be aligned with communities' experiences and needs. 
Therefore, despite the intent to promote social good, companies may fail to valorize the knowledge and expertise of harmed communities. 
As our workshop findings highlight, participants hold shared experiences and 
knowledge regarding the format of bias bounties that bounties would hinder them from providing. 

We synthesize critiques of bias bounties from queer communities into several salient themes (\Cref{fig:visual_results}) and find that they span all four components of the traditional AI development pipeline: problem formulation, data collection, algorithm development, and deployment. 
Bias bounties only allow for post-hoc interventions, providing limited options for feedback and control from queer communities. 
Because of this, we argue that harmed individuals must have the ability to self-actualize beyond the role of a `user,' `participant,' or `informant' of their own experienced harm; instead, communities must be offered the ability to co-design auditing processes and collaboratively generate knowledge throughout the AI pipeline, not just after a system is deployed. 
We argue for auditing research that enables transferring ownership of AI auditing processes to the communities that are harmed so that their experiences and knowledge may be integrated into new and more effective auditing methodologies. 
As future work, we encourage auditing researchers to explore feedback and control mechanisms in the context of other auditing processes and different marginalized communities, as well as concretely reimagine what community ownership of such processes would look like.

\begin{acks}
We thank the anonymous reviewers and Michael Madaio for their insightful and valuable feedback on this paper. We further thank the former Twitter bias bounty working group for early discussions about this project.
\end{acks}

\clearpage

\bibliographystyle{ACM-Reference-Format}
\bibliography{software}


\begin{thebibliography}{63}


\ifx \showCODEN    \undefined \def \showCODEN     #1{\unskip}     \fi
\ifx \showDOI      \undefined \def \showDOI       #1{#1}\fi
\ifx \showISBNx    \undefined \def \showISBNx     #1{\unskip}     \fi
\ifx \showISBNxiii \undefined \def \showISBNxiii  #1{\unskip}     \fi
\ifx \showISSN     \undefined \def \showISSN      #1{\unskip}     \fi
\ifx \showLCCN     \undefined \def \showLCCN      #1{\unskip}     \fi
\ifx \shownote     \undefined \def \shownote      #1{#1}          \fi
\ifx \showarticletitle \undefined \def \showarticletitle #1{#1}   \fi
\ifx \showURL      \undefined \def \showURL       {\relax}        \fi
\providecommand\bibfield[2]{#2}
\providecommand\bibinfo[2]{#2}
\providecommand\natexlab[1]{#1}
\providecommand\showeprint[2][]{arXiv:#2}

\bibitem[Agüera~y Arcas et~al\mbox{.}(2017)]%
        {newclothes}
\bibfield{author}{\bibinfo{person}{Blaise Agüera~y Arcas},
  \bibinfo{person}{Margaret Mitchell}, {and} \bibinfo{person}{Alexander
  Todorov}.} \bibinfo{year}{2017}\natexlab{}.
\newblock \bibinfo{title}{Physiognomy’s New Clothes}.
\newblock
\newblock
\urldef\tempurl%
\url{https://medium.com/@blaisea/physiognomys-new-clothes-f2d4b59fdd6a}
\showURL{%
\tempurl}


\bibitem[Bar~On(2018)]%
        {bar_on_2018}
\bibfield{author}{\bibinfo{person}{Amit~Elazari Bar~On}.}
  \bibinfo{year}{2018}\natexlab{}.
\newblock \bibinfo{title}{We need bug bounties for bad algorithms}.
\newblock
\newblock
\urldef\tempurl%
\url{https://www.vice.com/en/article/8xkyj3/we-need-bug-bounties-for-bad-algorithms}
\showURL{%
\tempurl}


\bibitem[Bellamy et~al\mbox{.}(2018)]%
        {aif360-oct-2018}
\bibfield{author}{\bibinfo{person}{Rachel K.~E. Bellamy},
  \bibinfo{person}{Kuntal Dey}, \bibinfo{person}{Michael Hind},
  \bibinfo{person}{Samuel~C. Hoffman}, \bibinfo{person}{Stephanie Houde},
  \bibinfo{person}{Kalapriya Kannan}, \bibinfo{person}{Pranay Lohia},
  \bibinfo{person}{Jacquelyn Martino}, \bibinfo{person}{Sameep Mehta},
  \bibinfo{person}{Aleksandra Mojsilovic}, \bibinfo{person}{Seema Nagar},
  \bibinfo{person}{Karthikeyan~Natesan Ramamurthy}, \bibinfo{person}{John
  Richards}, \bibinfo{person}{Diptikalyan Saha}, \bibinfo{person}{Prasanna
  Sattigeri}, \bibinfo{person}{Moninder Singh}, \bibinfo{person}{Kush~R.
  Varshney}, {and} \bibinfo{person}{Yunfeng Zhang}.}
  \bibinfo{year}{2018}\natexlab{}.
\newblock \bibinfo{title}{{AI Fairness} 360: An Extensible Toolkit for
  Detecting, Understanding, and Mitigating Unwanted Algorithmic Bias}.
\newblock
\newblock
\urldef\tempurl%
\url{https://arxiv.org/abs/1810.01943}
\showURL{%
\tempurl}


\bibitem[Bender and Friedman(2018)]%
        {bender-friedman-2018-data}
\bibfield{author}{\bibinfo{person}{Emily~M. Bender} {and}
  \bibinfo{person}{Batya Friedman}.} \bibinfo{year}{2018}\natexlab{}.
\newblock \showarticletitle{Data Statements for Natural Language Processing:
  Toward Mitigating System Bias and Enabling Better Science}.
\newblock \bibinfo{journal}{\emph{Transactions of the Association for
  Computational Linguistics}}  \bibinfo{volume}{6} (\bibinfo{year}{2018}),
  \bibinfo{pages}{587--604}.
\newblock
\urldef\tempurl%
\url{https://doi.org/10.1162/tacl_a_00041}
\showDOI{\tempurl}


\bibitem[Birhane et~al\mbox{.}(2022)]%
        {birhane2022power}
\bibfield{author}{\bibinfo{person}{Abeba Birhane}, \bibinfo{person}{William
  Isaac}, \bibinfo{person}{Vinodkumar Prabhakaran}, \bibinfo{person}{Mark
  Diaz}, \bibinfo{person}{Madeleine~Clare Elish}, \bibinfo{person}{Iason
  Gabriel}, {and} \bibinfo{person}{Shakir Mohamed}.}
  \bibinfo{year}{2022}\natexlab{}.
\newblock \showarticletitle{Power to the People? Opportunities and Challenges
  for Participatory AI}. In \bibinfo{booktitle}{\emph{Equity and Access in
  Algorithms, Mechanisms, and Optimization}} (Arlington, VA, USA)
  \emph{(\bibinfo{series}{EAAMO '22})}. \bibinfo{publisher}{Association for
  Computing Machinery}, \bibinfo{address}{New York, NY, USA}, Article
  \bibinfo{articleno}{6}, \bibinfo{numpages}{8}~pages.
\newblock
\showISBNx{9781450394772}
\urldef\tempurl%
\url{https://doi.org/10.1145/3551624.3555290}
\showDOI{\tempurl}


\bibitem[Blodgett et~al\mbox{.}(2021)]%
        {blodgett-etal-2021-stereotyping}
\bibfield{author}{\bibinfo{person}{Su~Lin Blodgett}, \bibinfo{person}{Gilsinia
  Lopez}, \bibinfo{person}{Alexandra Olteanu}, \bibinfo{person}{Robert Sim},
  {and} \bibinfo{person}{Hanna Wallach}.} \bibinfo{year}{2021}\natexlab{}.
\newblock \showarticletitle{Stereotyping {N}orwegian Salmon: An Inventory of
  Pitfalls in Fairness Benchmark Datasets}. In
  \bibinfo{booktitle}{\emph{Proceedings of the 59th Annual Meeting of the
  Association for Computational Linguistics and the 11th International Joint
  Conference on Natural Language Processing (Volume 1: Long Papers)}}.
  \bibinfo{publisher}{Association for Computational Linguistics},
  \bibinfo{address}{Online}, \bibinfo{pages}{1004--1015}.
\newblock
\urldef\tempurl%
\url{https://doi.org/10.18653/v1/2021.acl-long.81}
\showDOI{\tempurl}


\bibitem[Boorstein and Kelly(2022)]%
        {wapo_data}
\bibfield{author}{\bibinfo{person}{Michelle Boorstein} {and}
  \bibinfo{person}{Heather Kelly}.} \bibinfo{year}{2022}\natexlab{}.
\newblock \bibinfo{title}{Catholic group spent millions on app data that
  tracked gay priests}.
\newblock
\newblock
\urldef\tempurl%
\url{https://www.washingtonpost.com/dc-md-va/2023/03/09/catholics-gay-priests-grindr-data-bishops/}
\showURL{%
\tempurl}


\bibitem[Buccaneers(2023)]%
        {bias_buccaneers}
\bibfield{author}{\bibinfo{person}{Bias Buccaneers}.}
  \bibinfo{year}{2023}\natexlab{}.
\newblock \bibinfo{title}{Who are the Bias Buccaneers?}
\newblock
\newblock
\urldef\tempurl%
\url{https://www.biasbuccaneers.org/}
\showURL{%
\tempurl}


\bibitem[Carlisle(2021)]%
        {time_priest}
\bibfield{author}{\bibinfo{person}{Madeleine Carlisle}.}
  \bibinfo{year}{2021}\natexlab{}.
\newblock \bibinfo{title}{How the Alleged Outing of a Catholic Priest Shows the
  Sorry State of Data Privacy in America}.
\newblock
\newblock
\urldef\tempurl%
\url{https://time.com/6083323/bishop-pillar-grindr-data/}
\showURL{%
\tempurl}


\bibitem[Chowdhury and Williams(2021)]%
        {twitter-bias-bounty}
\bibfield{author}{\bibinfo{person}{Rumman Chowdhury} {and}
  \bibinfo{person}{Jutta Williams}.} \bibinfo{year}{2021}\natexlab{}.
\newblock \bibinfo{title}{Introducing Twitter’s first algorithmic bias bounty
  challenge}.
\newblock
\newblock
\urldef\tempurl%
\url{https://blog.twitter.com/engineering/en_us/topics/insights/2021/algorithmic-bias-bounty-challenge}
\showURL{%
\tempurl}


\bibitem[Clarke et~al\mbox{.}(2015)]%
        {clarke2015thematic}
\bibfield{author}{\bibinfo{person}{Victoria Clarke}, \bibinfo{person}{Virginia
  Braun}, {and} \bibinfo{person}{Nikki Hayfield}.}
  \bibinfo{year}{2015}\natexlab{}.
\newblock \showarticletitle{Thematic analysis}.
\newblock \bibinfo{journal}{\emph{Qualitative psychology: A practical guide to
  research methods}}  \bibinfo{volume}{3} (\bibinfo{year}{2015}),
  \bibinfo{pages}{222--248}.
\newblock


\bibitem[Cooke and Kothari(2001)]%
        {cooke2001participation}
\bibfield{author}{\bibinfo{person}{Bill Cooke} {and} \bibinfo{person}{Uma
  Kothari}.} \bibinfo{year}{2001}\natexlab{}.
\newblock \bibinfo{booktitle}{\emph{Participation}}.
\newblock \bibinfo{publisher}{Zed Books}, \bibinfo{address}{London, England}.
\newblock


\bibitem[Dalek et~al\mbox{.}(2021)]%
        {noaccess}
\bibfield{author}{\bibinfo{person}{Jakub Dalek}, \bibinfo{person}{Nica Dumlao},
  \bibinfo{person}{Miles Kenyon}, \bibinfo{person}{Irene Poetranto},
  \bibinfo{person}{Adam Senft}, \bibinfo{person}{Caroline Wesley},
  \bibinfo{person}{Arturo Filastò}, \bibinfo{person}{Maria Xynou}, {and}
  \bibinfo{person}{Amie Bishop}.} \bibinfo{year}{2021}\natexlab{}.
\newblock \bibinfo{title}{No Access: {LGBTIQ} Website Censorship in Six
  Countries}.
\newblock
\newblock
\urldef\tempurl%
\url{https://citizenlab.ca/2021/08/no-access-lgbtiq-website-censorship-in-six-countries/}
\showURL{%
\tempurl}


\bibitem[Dev et~al\mbox{.}(2021)]%
        {dev2021harms}
\bibfield{author}{\bibinfo{person}{Sunipa Dev}, \bibinfo{person}{Masoud
  Monajatipoor}, \bibinfo{person}{Anaelia Ovalle}, \bibinfo{person}{Arjun
  Subramonian}, \bibinfo{person}{Jeff Phillips}, {and} \bibinfo{person}{Kai-Wei
  Chang}.} \bibinfo{year}{2021}\natexlab{}.
\newblock \showarticletitle{Harms of Gender Exclusivity and Challenges in
  Non-Binary Representation in Language Technologies}. In
  \bibinfo{booktitle}{\emph{Proceedings of the 2021 Conference on Empirical
  Methods in Natural Language Processing}}. \bibinfo{publisher}{Association for
  Computational Linguistics}, \bibinfo{address}{Online and Punta Cana,
  Dominican Republic}, \bibinfo{pages}{1968--1994}.
\newblock
\urldef\tempurl%
\url{https://doi.org/10.18653/v1/2021.emnlp-main.150}
\showDOI{\tempurl}


\bibitem[DeVos et~al\mbox{.}(2022)]%
        {devos2022toward}
\bibfield{author}{\bibinfo{person}{Alicia DeVos}, \bibinfo{person}{Aditi
  Dhabalia}, \bibinfo{person}{Hong Shen}, \bibinfo{person}{Kenneth Holstein},
  {and} \bibinfo{person}{Motahhare Eslami}.} \bibinfo{year}{2022}\natexlab{}.
\newblock \showarticletitle{Toward User-Driven Algorithm Auditing:
  Investigating Users’ Strategies for Uncovering Harmful Algorithmic
  Behavior}. In \bibinfo{booktitle}{\emph{Proceedings of the 2022 CHI
  Conference on Human Factors in Computing Systems}} (New Orleans, LA, USA)
  \emph{(\bibinfo{series}{CHI '22})}. \bibinfo{publisher}{Association for
  Computing Machinery}, \bibinfo{address}{New York, NY, USA}, Article
  \bibinfo{articleno}{626}, \bibinfo{numpages}{19}~pages.
\newblock
\showISBNx{9781450391573}
\urldef\tempurl%
\url{https://doi.org/10.1145/3491102.3517441}
\showDOI{\tempurl}


\bibitem[Dias~Oliva et~al\mbox{.}(2021)]%
        {DiasOliva_Fighting_2021}
\bibfield{author}{\bibinfo{person}{Thiago Dias~Oliva},
  \bibinfo{person}{Dennys~Marcelo Antonialli}, {and}
  \bibinfo{person}{Alessandra Gomes}.} \bibinfo{year}{2021}\natexlab{}.
\newblock \showarticletitle{Fighting {{Hate Speech}}, {{Silencing Drag
  Queens}}? {{Artificial Intelligence}} in {{Content Moderation}} and {{Risks}}
  to {{LGBTQ Voices Online}}}.
\newblock \bibinfo{journal}{\emph{Sexuality \& Culture}} \bibinfo{volume}{25},
  \bibinfo{number}{2} (\bibinfo{date}{April} \bibinfo{year}{2021}),
  \bibinfo{pages}{700--732}.
\newblock
\showISSN{1095-5143, 1936-4822}
\urldef\tempurl%
\url{http://link.springer.com/10.1007/s12119-020-09790-w}
\showURL{%
\tempurl}


\bibitem[D'Ignazio and Klein(2023)]%
        {d2020data}
\bibfield{author}{\bibinfo{person}{Catherine D'Ignazio} {and}
  \bibinfo{person}{Lauren~F Klein}.} \bibinfo{year}{2023}\natexlab{}.
\newblock \bibinfo{booktitle}{\emph{Data feminism}}.
\newblock \bibinfo{publisher}{MIT Press}, \bibinfo{address}{London, England}.
\newblock


\bibitem[Dodge et~al\mbox{.}(2021)]%
        {dodge2021documenting}
\bibfield{author}{\bibinfo{person}{Jesse Dodge}, \bibinfo{person}{Maarten Sap},
  \bibinfo{person}{Ana Marasovi{\'c}}, \bibinfo{person}{William Agnew},
  \bibinfo{person}{Gabriel Ilharco}, \bibinfo{person}{Dirk Groeneveld},
  \bibinfo{person}{Margaret Mitchell}, {and} \bibinfo{person}{Matt Gardner}.}
  \bibinfo{year}{2021}\natexlab{}.
\newblock \showarticletitle{Documenting Large Webtext Corpora: A Case Study on
  the Colossal Clean Crawled Corpus}. In \bibinfo{booktitle}{\emph{Proceedings
  of the 2021 Conference on Empirical Methods in Natural Language Processing}}.
  \bibinfo{publisher}{Association for Computational Linguistics},
  \bibinfo{address}{Online and Punta Cana, Dominican Republic},
  \bibinfo{pages}{1286--1305}.
\newblock
\urldef\tempurl%
\url{https://doi.org/10.18653/v1/2021.emnlp-main.98}
\showDOI{\tempurl}


\bibitem[dungeon(2019)]%
        {ai_dungeon_2019}
\bibfield{author}{\bibinfo{person}{AI dungeon}.}
  \bibinfo{year}{2019}\natexlab{}.
\newblock \bibinfo{title}{AI dungeon}.
\newblock
\newblock
\urldef\tempurl%
\url{https://play.aidungeon.io/main/home}
\showURL{%
\tempurl}


\bibitem[Embury-Dennis(2020)]%
        {bulliedblackmailed}
\bibfield{author}{\bibinfo{person}{Tom Embury-Dennis}.}
  \bibinfo{year}{2020}\natexlab{}.
\newblock \bibinfo{title}{Bullied and blackmailed: Gay men in Morocco falling
  victims to outing campaign sparked by Instagram model}.
\newblock
\newblock
\urldef\tempurl%
\url{https://www.independent.co.uk/news/world/africa/gay-men-morocco-dating-apps-grindr-instagram-sofia-taloni-a9486386.html}
\showURL{%
\tempurl}


\bibitem[Felkner et~al\mbox{.}(2023)]%
        {Felkner2022TowardsWD}
\bibfield{author}{\bibinfo{person}{Virginia Felkner},
  \bibinfo{person}{Ho-Chun~Herbert Chang}, \bibinfo{person}{Eugene Jang}, {and}
  \bibinfo{person}{Jonathan May}.} \bibinfo{year}{2023}\natexlab{}.
\newblock \showarticletitle{{W}ino{Q}ueer: A Community-in-the-Loop Benchmark
  for Anti-{LGBTQ}+ Bias in Large Language Models}. In
  \bibinfo{booktitle}{\emph{Proceedings of the 61st Annual Meeting of the
  Association for Computational Linguistics (Volume 1: Long Papers)}}.
  \bibinfo{publisher}{Association for Computational Linguistics},
  \bibinfo{address}{Toronto, Canada}, \bibinfo{pages}{9126--9140}.
\newblock
\urldef\tempurl%
\url{https://aclanthology.org/2023.acl-long.507}
\showURL{%
\tempurl}


\bibitem[Fine and Torre(2006)]%
        {fine2006intimate}
\bibfield{author}{\bibinfo{person}{Michelle Fine} {and}
  \bibinfo{person}{Mar{\'\i}a~Elena Torre}.} \bibinfo{year}{2006}\natexlab{}.
\newblock \showarticletitle{Intimate details: Participatory action research in
  prison}.
\newblock \bibinfo{journal}{\emph{Action Research}} \bibinfo{volume}{4},
  \bibinfo{number}{3} (\bibinfo{year}{2006}), \bibinfo{pages}{253--269}.
\newblock


\bibitem[Fox(2020)]%
        {independant_grindr}
\bibfield{author}{\bibinfo{person}{Gemma Fox}.}
  \bibinfo{year}{2020}\natexlab{}.
\newblock \bibinfo{title}{Egypt police ‘using dating apps’ to find and
  imprison LGBT+ people}.
\newblock
\newblock
\urldef\tempurl%
\url{https://www.independent.co.uk/news/world/middle-east/egypt-lgbt-gay-facebook-grindr-jail-torture-police-hrw-b742231.html}
\showURL{%
\tempurl}


\bibitem[Gardner et~al\mbox{.}(2018)]%
        {Gardner2017AllenNLP}
\bibfield{author}{\bibinfo{person}{Matt Gardner}, \bibinfo{person}{Joel Grus},
  \bibinfo{person}{Mark Neumann}, \bibinfo{person}{Oyvind Tafjord},
  \bibinfo{person}{Pradeep Dasigi}, \bibinfo{person}{Nelson~F. Liu},
  \bibinfo{person}{Matthew Peters}, \bibinfo{person}{Michael Schmitz}, {and}
  \bibinfo{person}{Luke Zettlemoyer}.} \bibinfo{year}{2018}\natexlab{}.
\newblock \showarticletitle{{A}llen{NLP}: A Deep Semantic Natural Language
  Processing Platform}. In \bibinfo{booktitle}{\emph{Proceedings of Workshop
  for {NLP} Open Source Software ({NLP}-{OSS})}}.
  \bibinfo{publisher}{Association for Computational Linguistics},
  \bibinfo{address}{Melbourne, Australia}, \bibinfo{pages}{1--6}.
\newblock
\urldef\tempurl%
\url{https://doi.org/10.18653/v1/W18-2501}
\showDOI{\tempurl}


\bibitem[Gebru et~al\mbox{.}(2018)]%
        {Gebru2018DatasheetsFD}
\bibfield{author}{\bibinfo{person}{Timnit Gebru}, \bibinfo{person}{Jamie~H.
  Morgenstern}, \bibinfo{person}{Briana Vecchione},
  \bibinfo{person}{Jennifer~Wortman Vaughan}, \bibinfo{person}{Hanna~M.
  Wallach}, \bibinfo{person}{Hal Daum{\'e}}, {and} \bibinfo{person}{Kate
  Crawford}.} \bibinfo{year}{2018}\natexlab{}.
\newblock \showarticletitle{Datasheets for datasets}.
\newblock \bibinfo{journal}{\emph{Commun. ACM}}  \bibinfo{volume}{64}
  (\bibinfo{year}{2018}), \bibinfo{pages}{86--92}.
\newblock


\bibitem[Glesne(1989)]%
        {glesne1989rapport}
\bibfield{author}{\bibinfo{person}{Corrine Glesne}.}
  \bibinfo{year}{1989}\natexlab{}.
\newblock \showarticletitle{Rapport and friendship in ethnographic research}.
\newblock \bibinfo{journal}{\emph{Internation Journal of Qualitative Studies in
  Education}} \bibinfo{volume}{2}, \bibinfo{number}{1} (\bibinfo{year}{1989}),
  \bibinfo{pages}{45--54}.
\newblock


\bibitem[Globus-Harris et~al\mbox{.}(2022)]%
        {globus2022algorithmic}
\bibfield{author}{\bibinfo{person}{Ira Globus-Harris}, \bibinfo{person}{Michael
  Kearns}, {and} \bibinfo{person}{Aaron Roth}.}
  \bibinfo{year}{2022}\natexlab{}.
\newblock \showarticletitle{An Algorithmic Framework for Bias Bounties}. In
  \bibinfo{booktitle}{\emph{Proceedings of the 2022 ACM Conference on Fairness,
  Accountability, and Transparency}} (Seoul, Republic of Korea)
  \emph{(\bibinfo{series}{FAccT '22})}. \bibinfo{publisher}{Association for
  Computing Machinery}, \bibinfo{address}{New York, NY, USA},
  \bibinfo{pages}{1106–1124}.
\newblock
\showISBNx{9781450393522}
\urldef\tempurl%
\url{https://doi.org/10.1145/3531146.3533172}
\showDOI{\tempurl}


\bibitem[Gray and Suri(2019)]%
        {gray2019ghost}
\bibfield{author}{\bibinfo{person}{Mary~L Gray} {and}
  \bibinfo{person}{Siddharth Suri}.} \bibinfo{year}{2019}\natexlab{}.
\newblock \bibinfo{booktitle}{\emph{Ghost work}}.
\newblock \bibinfo{publisher}{Houghton Mifflin Harcourt Publishing Company},
  \bibinfo{address}{Boston, MA}.
\newblock


\bibitem[Green et~al\mbox{.}(2003)]%
        {green2003appendix}
\bibfield{author}{\bibinfo{person}{LW Green}, \bibinfo{person}{MA George},
  {et~al\mbox{.}}} \bibinfo{year}{2003}\natexlab{}.
\newblock \showarticletitle{Appendix C: Guidelines for participatory research
  in health promotion}.
\newblock In \bibinfo{booktitle}{\emph{Community-based participatory research
  for health}}, \bibfield{editor}{\bibinfo{person}{M.~Minkler} {and}
  \bibinfo{person}{N.~Wallerstein}} (Eds.). \bibinfo{publisher}{Jossey-Bass},
  \bibinfo{address}{San Francisco, CA}.
\newblock


\bibitem[Groves et~al\mbox{.}(2023)]%
        {Groves2023GoingPT}
\bibfield{author}{\bibinfo{person}{Lara Groves}, \bibinfo{person}{Aidan
  Peppin}, \bibinfo{person}{Andrew Strait}, {and} \bibinfo{person}{Jenny
  Brennan}.} \bibinfo{year}{2023}\natexlab{}.
\newblock \showarticletitle{Going Public: The Role of Public Participation
  Approaches in Commercial AI Labs}. In \bibinfo{booktitle}{\emph{Proceedings
  of the 2023 ACM Conference on Fairness, Accountability, and Transparency}}
  (Chicago, IL, USA) \emph{(\bibinfo{series}{FAccT '23})}.
  \bibinfo{publisher}{Association for Computing Machinery},
  \bibinfo{address}{New York, NY, USA}, \bibinfo{pages}{1162–1173}.
\newblock
\showISBNx{9798400701924}
\urldef\tempurl%
\url{https://doi.org/10.1145/3593013.3594071}
\showDOI{\tempurl}


\bibitem[Hacker and Taylor(2011)]%
        {hacker-taylor}
\bibfield{author}{\bibinfo{person}{Karen Hacker} {and}
  \bibinfo{person}{J.~Glover Taylor}.} \bibinfo{year}{2011}\natexlab{}.
\newblock \bibinfo{title}{Community-Engaged Research 101}.
\newblock
\newblock
\urldef\tempurl%
\url{https://catalyst.harvard.edu/publications-documents/community-engaged-research-101-2/}
\showURL{%
\tempurl}


\bibitem[Hamidi et~al\mbox{.}(2020)]%
        {hamidi2020gender}
\bibfield{author}{\bibinfo{person}{Foad Hamidi}, \bibinfo{person}{Morgan
  Scheuerman}, {and} \bibinfo{person}{Stacy Branham}.}
  \bibinfo{year}{2020}\natexlab{}.
\newblock \bibinfo{title}{Gender is personal--not computational}.
\newblock
\newblock


\bibitem[Im et~al\mbox{.}(2021)]%
        {im2021yes}
\bibfield{author}{\bibinfo{person}{Jane Im}, \bibinfo{person}{Jill Dimond},
  \bibinfo{person}{Melody Berton}, \bibinfo{person}{Una Lee},
  \bibinfo{person}{Katherine Mustelier}, \bibinfo{person}{Mark~S. Ackerman},
  {and} \bibinfo{person}{Eric Gilbert}.} \bibinfo{year}{2021}\natexlab{}.
\newblock \showarticletitle{Yes: Affirmative Consent as a Theoretical Framework
  for Understanding and Imagining Social Platforms}. In
  \bibinfo{booktitle}{\emph{Proceedings of the 2021 CHI Conference on Human
  Factors in Computing Systems}} (Yokohama, Japan) \emph{(\bibinfo{series}{CHI
  '21})}. \bibinfo{publisher}{Association for Computing Machinery},
  \bibinfo{address}{New York, NY, USA}, Article \bibinfo{articleno}{403},
  \bibinfo{numpages}{18}~pages.
\newblock
\showISBNx{9781450380966}
\urldef\tempurl%
\url{https://doi.org/10.1145/3411764.3445778}
\showDOI{\tempurl}


\bibitem[Kayser-Bril(2021)]%
        {Kayser-Bril}
\bibfield{author}{\bibinfo{person}{Nicolas Kayser-Bril}.}
  \bibinfo{year}{2021}\natexlab{}.
\newblock \bibinfo{title}{Twitter’s algorithmic bias bug bounty could be the
  way forward, if regulators step in}.
\newblock
\newblock
\urldef\tempurl%
\url{https://algorithmwatch.org/en/twitters-algorithmic-bias-bug-bounty/}
\showURL{%
\tempurl}


\bibitem[Kenway et~al\mbox{.}(2022)]%
        {kenway_francois_costanza-chock_raji_buolamwini}
\bibfield{author}{\bibinfo{person}{Josh Kenway}, \bibinfo{person}{Camille
  François}, \bibinfo{person}{Sasha Costanza-Chock},
  \bibinfo{person}{Inioluwa~Deborah Raji}, {and} \bibinfo{person}{Joy
  Buolamwini}.} \bibinfo{year}{2022}\natexlab{}.
\newblock \bibinfo{title}{Bug bounties for algorithmic harms?}
\newblock
\newblock
\urldef\tempurl%
\url{https://www.ajl.org/bugs}
\showURL{%
\tempurl}


\bibitem[Keyes(2019)]%
        {keyes2019counting}
\bibfield{author}{\bibinfo{person}{Os Keyes}.} \bibinfo{year}{2019}\natexlab{}.
\newblock \bibinfo{title}{Counting the Countless: Why data science is a
  profound threat for queer people}.
\newblock
\newblock


\bibitem[Keyes and Austin(2022)]%
        {keyes2022feeling}
\bibfield{author}{\bibinfo{person}{Os Keyes} {and} \bibinfo{person}{Jeanie
  Austin}.} \bibinfo{year}{2022}\natexlab{}.
\newblock \showarticletitle{Feeling fixes: Mess and emotion in algorithmic
  audits}.
\newblock \bibinfo{journal}{\emph{Big Data \& Society}} \bibinfo{volume}{9},
  \bibinfo{number}{2} (\bibinfo{year}{2022}),
  \bibinfo{pages}{20539517221113772}.
\newblock


\bibitem[Kormilitzin et~al\mbox{.}(2023)]%
        {kormilitzin2023participatory}
\bibfield{author}{\bibinfo{person}{Andrey Kormilitzin}, \bibinfo{person}{Nenad
  Tomasev}, \bibinfo{person}{Kevin~R. McKee}, {and} \bibinfo{person}{Dan~W.
  Joyce}.} \bibinfo{year}{2023}\natexlab{}.
\newblock \showarticletitle{A participatory initiative to include {LGBT}+
  voices in {AI} for mental health}.
\newblock \bibinfo{journal}{\emph{Nature Medicine}} \bibinfo{volume}{29},
  \bibinfo{number}{1} (\bibinfo{date}{Jan.} \bibinfo{year}{2023}),
  \bibinfo{pages}{10--11}.
\newblock
\urldef\tempurl%
\url{https://doi.org/10.1038/s41591-022-02137-y}
\showDOI{\tempurl}


\bibitem[Li et~al\mbox{.}(2023)]%
        {li2023userdriven}
\bibfield{author}{\bibinfo{person}{Rena Li}, \bibinfo{person}{Sara Kingsley},
  \bibinfo{person}{Chelsea Fan}, \bibinfo{person}{Proteeti Sinha},
  \bibinfo{person}{Nora Wai}, \bibinfo{person}{Jaimie Lee},
  \bibinfo{person}{Hong Shen}, \bibinfo{person}{Motahhare Eslami}, {and}
  \bibinfo{person}{Jason Hong}.} \bibinfo{year}{2023}\natexlab{}.
\newblock \showarticletitle{Participation and Division of Labor in User-Driven
  Algorithm Audits: How Do Everyday Users Work Together to Surface Algorithmic
  Harms?}. In \bibinfo{booktitle}{\emph{Proceedings of the 2023 CHI Conference
  on Human Factors in Computing Systems}} (Hamburg, Germany)
  \emph{(\bibinfo{series}{CHI '23})}. \bibinfo{publisher}{Association for
  Computing Machinery}, \bibinfo{address}{New York, NY, USA}, Article
  \bibinfo{articleno}{567}, \bibinfo{numpages}{19}~pages.
\newblock
\showISBNx{9781450394215}
\urldef\tempurl%
\url{https://doi.org/10.1145/3544548.3582074}
\showDOI{\tempurl}


\bibitem[Lu et~al\mbox{.}(2022)]%
        {lu2022subverting}
\bibfield{author}{\bibinfo{person}{Christina Lu}, \bibinfo{person}{Jackie Kay},
  {and} \bibinfo{person}{Kevin McKee}.} \bibinfo{year}{2022}\natexlab{}.
\newblock \showarticletitle{Subverting Machines, Fluctuating Identities:
  Re-Learning Human Categorization}. In \bibinfo{booktitle}{\emph{Proceedings
  of the 2022 ACM Conference on Fairness, Accountability, and Transparency}}
  (Seoul, Republic of Korea) \emph{(\bibinfo{series}{FAccT '22})}.
  \bibinfo{publisher}{Association for Computing Machinery},
  \bibinfo{address}{New York, NY, USA}, \bibinfo{pages}{1005–1015}.
\newblock
\showISBNx{9781450393522}
\urldef\tempurl%
\url{https://doi.org/10.1145/3531146.3533161}
\showDOI{\tempurl}


\bibitem[Metcalf et~al\mbox{.}(2021)]%
        {Metcalf2020AlgorithmicIA}
\bibfield{author}{\bibinfo{person}{Jacob Metcalf}, \bibinfo{person}{Emanuel
  Moss}, \bibinfo{person}{Elizabeth~Anne Watkins}, \bibinfo{person}{Ranjit
  Singh}, {and} \bibinfo{person}{Madeleine~Clare Elish}.}
  \bibinfo{year}{2021}\natexlab{}.
\newblock \showarticletitle{Algorithmic Impact Assessments and Accountability:
  The Co-Construction of Impacts}. In \bibinfo{booktitle}{\emph{Proceedings of
  the 2021 ACM Conference on Fairness, Accountability, and Transparency}}
  (Virtual Event, Canada) \emph{(\bibinfo{series}{FAccT '21})}.
  \bibinfo{publisher}{Association for Computing Machinery},
  \bibinfo{address}{New York, NY, USA}, \bibinfo{pages}{735–746}.
\newblock
\showISBNx{9781450383097}
\urldef\tempurl%
\url{https://doi.org/10.1145/3442188.3445935}
\showDOI{\tempurl}


\bibitem[Mitchell et~al\mbox{.}(2019)]%
        {Mitchell2018ModelCF}
\bibfield{author}{\bibinfo{person}{Margaret Mitchell}, \bibinfo{person}{Simone
  Wu}, \bibinfo{person}{Andrew Zaldivar}, \bibinfo{person}{Parker Barnes},
  \bibinfo{person}{Lucy Vasserman}, \bibinfo{person}{Ben Hutchinson},
  \bibinfo{person}{Elena Spitzer}, \bibinfo{person}{Inioluwa~Deborah Raji},
  {and} \bibinfo{person}{Timnit Gebru}.} \bibinfo{year}{2019}\natexlab{}.
\newblock \showarticletitle{Model Cards for Model Reporting}. In
  \bibinfo{booktitle}{\emph{Proceedings of the Conference on Fairness,
  Accountability, and Transparency}} (Atlanta, GA, USA)
  \emph{(\bibinfo{series}{FAT* '19})}. \bibinfo{publisher}{Association for
  Computing Machinery}, \bibinfo{address}{New York, NY, USA},
  \bibinfo{pages}{220–229}.
\newblock
\showISBNx{9781450361255}
\urldef\tempurl%
\url{https://doi.org/10.1145/3287560.3287596}
\showDOI{\tempurl}


\bibitem[Monea(2022)]%
        {monea2022digital}
\bibfield{author}{\bibinfo{person}{Alexander Monea}.}
  \bibinfo{year}{2022}\natexlab{}.
\newblock \bibinfo{booktitle}{\emph{The digital closet}}.
\newblock \bibinfo{publisher}{MIT Press}, \bibinfo{address}{Cambridge, MA}.
\newblock


\bibitem[Nichol et~al\mbox{.}(2021)]%
        {nichol2021glide}
\bibfield{author}{\bibinfo{person}{Alex Nichol}, \bibinfo{person}{Prafulla
  Dhariwal}, \bibinfo{person}{Aditya Ramesh}, \bibinfo{person}{Pranav Shyam},
  \bibinfo{person}{Pamela Mishkin}, \bibinfo{person}{Bob McGrew},
  \bibinfo{person}{Ilya Sutskever}, {and} \bibinfo{person}{Mark Chen}.}
  \bibinfo{year}{2021}\natexlab{}.
\newblock \bibinfo{title}{Glide: Towards photorealistic image generation and
  editing with text-guided diffusion models}.
\newblock
\newblock


\bibitem[Onuoha and Nucera(2018)]%
        {peoples}
\bibfield{author}{\bibinfo{person}{Mimi Onuoha} {and} \bibinfo{person}{Nucera
  Nucera}.} \bibinfo{year}{2018}\natexlab{}.
\newblock \bibinfo{booktitle}{\emph{The People's Guide to Artificial
  Intelligence}}.
\newblock \bibinfo{publisher}{Allied Media Projects},
  \bibinfo{address}{virtual}.
\newblock


\bibitem[Ovalle et~al\mbox{.}(2023)]%
        {ovalle2023m}
\bibfield{author}{\bibinfo{person}{Anaelia Ovalle}, \bibinfo{person}{Palash
  Goyal}, \bibinfo{person}{Jwala Dhamala}, \bibinfo{person}{Zachary Jaggers},
  \bibinfo{person}{Kai-Wei Chang}, \bibinfo{person}{Aram Galstyan},
  \bibinfo{person}{Richard Zemel}, {and} \bibinfo{person}{Rahul Gupta}.}
  \bibinfo{year}{2023}\natexlab{}.
\newblock \showarticletitle{“I’m Fully Who I Am”: Towards Centering
  Transgender and Non-Binary Voices to Measure Biases in Open Language
  Generation}. In \bibinfo{booktitle}{\emph{Proceedings of the 2023 ACM
  Conference on Fairness, Accountability, and Transparency}} (Chicago, IL, USA)
  \emph{(\bibinfo{series}{FAccT '23})}. \bibinfo{publisher}{Association for
  Computing Machinery}, \bibinfo{address}{New York, NY, USA},
  \bibinfo{pages}{1246–1266}.
\newblock
\showISBNx{9798400701924}
\urldef\tempurl%
\url{https://doi.org/10.1145/3593013.3594078}
\showDOI{\tempurl}


\bibitem[Powell et~al\mbox{.}(2020)]%
        {powell2020digital}
\bibfield{author}{\bibinfo{person}{Anastasia Powell}, \bibinfo{person}{Adrian~J
  Scott}, {and} \bibinfo{person}{Nicola Henry}.}
  \bibinfo{year}{2020}\natexlab{}.
\newblock \showarticletitle{Digital harassment and abuse: Experiences of
  sexuality and gender minority adults}.
\newblock \bibinfo{journal}{\emph{European journal of criminology}}
  \bibinfo{volume}{17}, \bibinfo{number}{2} (\bibinfo{year}{2020}),
  \bibinfo{pages}{199--223}.
\newblock


\bibitem[Queerinai et~al\mbox{.}(2023)]%
        {qai2023}
\bibfield{author}{\bibinfo{person}{Organizers~Of Queerinai},
  \bibinfo{person}{Anaelia Ovalle}, \bibinfo{person}{Arjun Subramonian},
  \bibinfo{person}{Ashwin Singh}, \bibinfo{person}{Claas Voelcker},
  \bibinfo{person}{Danica~J. Sutherland}, \bibinfo{person}{Davide Locatelli},
  \bibinfo{person}{Eva Breznik}, \bibinfo{person}{Filip Klubicka},
  \bibinfo{person}{Hang Yuan}, \bibinfo{person}{Hetvi J}, \bibinfo{person}{Huan
  Zhang}, \bibinfo{person}{Jaidev Shriram}, \bibinfo{person}{Kruno Lehman},
  \bibinfo{person}{Luca Soldaini}, \bibinfo{person}{Maarten Sap},
  \bibinfo{person}{Marc~Peter Deisenroth}, \bibinfo{person}{Maria~Leonor
  Pacheco}, \bibinfo{person}{Maria Ryskina}, \bibinfo{person}{Martin Mundt},
  \bibinfo{person}{Milind Agarwal}, \bibinfo{person}{Nyx Mclean},
  \bibinfo{person}{Pan Xu}, \bibinfo{person}{A Pranav}, \bibinfo{person}{Raj
  Korpan}, \bibinfo{person}{Ruchira Ray}, \bibinfo{person}{Sarah Mathew},
  \bibinfo{person}{Sarthak Arora}, \bibinfo{person}{St John},
  \bibinfo{person}{Tanvi Anand}, \bibinfo{person}{Vishakha Agrawal},
  \bibinfo{person}{William Agnew}, \bibinfo{person}{Yanan Long},
  \bibinfo{person}{Zijie~J. Wang}, \bibinfo{person}{Zeerak Talat},
  \bibinfo{person}{Avijit Ghosh}, \bibinfo{person}{Nathaniel Dennler},
  \bibinfo{person}{Michael Noseworthy}, \bibinfo{person}{Sharvani Jha},
  \bibinfo{person}{Emi Baylor}, \bibinfo{person}{Aditya Joshi},
  \bibinfo{person}{Natalia~Y. Bilenko}, \bibinfo{person}{Andrew Mcnamara},
  \bibinfo{person}{Raphael Gontijo-Lopes}, \bibinfo{person}{Alex Markham},
  \bibinfo{person}{Evyn Dong}, \bibinfo{person}{Jackie Kay},
  \bibinfo{person}{Manu Saraswat}, \bibinfo{person}{Nikhil Vytla}, {and}
  \bibinfo{person}{Luke Stark}.} \bibinfo{year}{2023}\natexlab{}.
\newblock \showarticletitle{Queer In AI: A Case Study in Community-Led
  Participatory AI}. In \bibinfo{booktitle}{\emph{Proceedings of the 2023 ACM
  Conference on Fairness, Accountability, and Transparency}} (Chicago, IL, USA)
  \emph{(\bibinfo{series}{FAccT '23})}. \bibinfo{publisher}{Association for
  Computing Machinery}, \bibinfo{address}{New York, NY, USA},
  \bibinfo{pages}{1882–1895}.
\newblock
\showISBNx{9798400701924}
\urldef\tempurl%
\url{https://doi.org/10.1145/3593013.3594134}
\showDOI{\tempurl}


\bibitem[QueerInAI et~al\mbox{.}(2021)]%
        {agnew2021rebuilding}
\bibfield{author}{\bibinfo{person}{Organizers~of QueerInAI},
  \bibinfo{person}{Ashwin S}, \bibinfo{person}{William Agnew},
  \bibinfo{person}{Hetvi Jethwani}, {and} \bibinfo{person}{Arjun Subramonian}.}
  \bibinfo{year}{2021}\natexlab{}.
\newblock \bibinfo{title}{Rebuilding Trust: Queer in AI Approach to Artificial
  Intelligence Risk Management}.
\newblock
\newblock
\urldef\tempurl%
\url{queerinai.org/risk-management}
\showURL{%
\tempurl}


\bibitem[Raji and Buolamwini(2019)]%
        {Raji2019Actionable}
\bibfield{author}{\bibinfo{person}{Inioluwa~Deborah Raji} {and}
  \bibinfo{person}{Joy Buolamwini}.} \bibinfo{year}{2019}\natexlab{}.
\newblock \showarticletitle{Actionable Auditing: Investigating the Impact of
  Publicly Naming Biased Performance Results of Commercial AI Products}. In
  \bibinfo{booktitle}{\emph{Proceedings of the 2019 AAAI/ACM Conference on AI,
  Ethics, and Society}} (Honolulu, HI, USA) \emph{(\bibinfo{series}{AIES
  '19})}. \bibinfo{publisher}{Association for Computing Machinery},
  \bibinfo{address}{New York, NY, USA}, \bibinfo{pages}{429–435}.
\newblock
\showISBNx{9781450363242}
\urldef\tempurl%
\url{https://doi.org/10.1145/3306618.3314244}
\showDOI{\tempurl}


\bibitem[Raji et~al\mbox{.}(2022)]%
        {Raji2022Outsider}
\bibfield{author}{\bibinfo{person}{Inioluwa~Deborah Raji},
  \bibinfo{person}{Peggy Xu}, \bibinfo{person}{Colleen Honigsberg}, {and}
  \bibinfo{person}{Daniel Ho}.} \bibinfo{year}{2022}\natexlab{}.
\newblock \showarticletitle{Outsider Oversight: Designing a Third Party Audit
  Ecosystem for AI Governance}. In \bibinfo{booktitle}{\emph{Proceedings of the
  2022 AAAI/ACM Conference on AI, Ethics, and Society}} (Oxford, United
  Kingdom) \emph{(\bibinfo{series}{AIES '22})}. \bibinfo{publisher}{Association
  for Computing Machinery}, \bibinfo{address}{New York, NY, USA},
  \bibinfo{pages}{557–571}.
\newblock
\showISBNx{9781450392471}
\urldef\tempurl%
\url{https://doi.org/10.1145/3514094.3534181}
\showDOI{\tempurl}


\bibitem[Ramesh et~al\mbox{.}(2021)]%
        {ramesh2021zero}
\bibfield{author}{\bibinfo{person}{Aditya Ramesh}, \bibinfo{person}{Mikhail
  Pavlov}, \bibinfo{person}{Gabriel Goh}, \bibinfo{person}{Scott Gray},
  \bibinfo{person}{Chelsea Voss}, \bibinfo{person}{Alec Radford},
  \bibinfo{person}{Mark Chen}, {and} \bibinfo{person}{Ilya Sutskever}.}
  \bibinfo{year}{2021}\natexlab{}.
\newblock \showarticletitle{Zero-Shot Text-to-Image Generation}. In
  \bibinfo{booktitle}{\emph{Proceedings of the 38th International Conference on
  Machine Learning}} \emph{(\bibinfo{series}{Proceedings of Machine Learning
  Research}, Vol.~\bibinfo{volume}{139})},
  \bibfield{editor}{\bibinfo{person}{Marina Meila} {and} \bibinfo{person}{Tong
  Zhang}} (Eds.). \bibinfo{publisher}{PMLR}, \bibinfo{address}{virtual},
  \bibinfo{pages}{8821--8831}.
\newblock
\urldef\tempurl%
\url{https://proceedings.mlr.press/v139/ramesh21a.html}
\showURL{%
\tempurl}


\bibitem[Rando et~al\mbox{.}(2022)]%
        {rando2022red}
\bibfield{author}{\bibinfo{person}{Javier Rando}, \bibinfo{person}{Daniel
  Paleka}, \bibinfo{person}{David Lindner}, \bibinfo{person}{Lennard Heim},
  {and} \bibinfo{person}{Florian Tram{\`e}r}.} \bibinfo{year}{2022}\natexlab{}.
\newblock \bibinfo{title}{Red-Teaming the Stable Diffusion Safety Filter}.
\newblock
\newblock


\bibitem[Saleiro et~al\mbox{.}(2018)]%
        {2018aequitas}
\bibfield{author}{\bibinfo{person}{Pedro Saleiro}, \bibinfo{person}{Benedict
  Kuester}, \bibinfo{person}{Abby Stevens}, \bibinfo{person}{Ari Anisfeld},
  \bibinfo{person}{Loren Hinkson}, \bibinfo{person}{Jesse London}, {and}
  \bibinfo{person}{Rayid Ghani}.} \bibinfo{year}{2018}\natexlab{}.
\newblock \bibinfo{title}{Aequitas: A Bias and Fairness Audit Toolkit}.
\newblock
\newblock


\bibitem[Shen et~al\mbox{.}(2021)]%
        {hong2021everyday}
\bibfield{author}{\bibinfo{person}{Hong Shen}, \bibinfo{person}{Alicia DeVos},
  \bibinfo{person}{Motahhare Eslami}, {and} \bibinfo{person}{Kenneth
  Holstein}.} \bibinfo{year}{2021}\natexlab{}.
\newblock \showarticletitle{Everyday Algorithm Auditing: Understanding the
  Power of Everyday Users in Surfacing Harmful Algorithmic Behaviors}.
\newblock \bibinfo{journal}{\emph{Proc. ACM Hum.-Comput. Interact.}}
  \bibinfo{volume}{5}, \bibinfo{number}{CSCW2}, Article
  \bibinfo{articleno}{433} (\bibinfo{date}{oct} \bibinfo{year}{2021}),
  \bibinfo{numpages}{29}~pages.
\newblock
\urldef\tempurl%
\url{https://doi.org/10.1145/3479577}
\showDOI{\tempurl}


\bibitem[Sloane et~al\mbox{.}(2022)]%
        {sloane22participation}
\bibfield{author}{\bibinfo{person}{Mona Sloane}, \bibinfo{person}{Emanuel
  Moss}, \bibinfo{person}{Olaitan Awomolo}, {and} \bibinfo{person}{Laura
  Forlano}.} \bibinfo{year}{2022}\natexlab{}.
\newblock \showarticletitle{Participation Is Not a Design Fix for Machine
  Learning}. In \bibinfo{booktitle}{\emph{Equity and Access in Algorithms,
  Mechanisms, and Optimization}} (Arlington, VA, USA)
  \emph{(\bibinfo{series}{EAAMO '22})}. \bibinfo{publisher}{Association for
  Computing Machinery}, \bibinfo{address}{New York, NY, USA}, Article
  \bibinfo{articleno}{1}, \bibinfo{numpages}{6}~pages.
\newblock
\showISBNx{9781450394772}
\urldef\tempurl%
\url{https://doi.org/10.1145/3551624.3555285}
\showDOI{\tempurl}


\bibitem[Smith et~al\mbox{.}(2021)]%
        {salty}
\bibfield{author}{\bibinfo{person}{Shakira Smith}, \bibinfo{person}{Oliver~L
  Haimson}, \bibinfo{person}{Claire Fitzsimmons}, {and}
  \bibinfo{person}{Nikki~Echarte Brown}.} \bibinfo{year}{2021}\natexlab{}.
\newblock \bibinfo{title}{Censorship of Marginalized Communities on Instagram}.
\newblock
\newblock
\urldef\tempurl%
\url{https://saltyworld.net/exclusive-report-censorship-of-marginalized-communities-on-instagram-2021-pdf-download/}
\showURL{%
\tempurl}


\bibitem[Spade(2020)]%
        {spade2020solidarity}
\bibfield{author}{\bibinfo{person}{Dean Spade}.}
  \bibinfo{year}{2020}\natexlab{}.
\newblock \showarticletitle{Solidarity not charity: Mutual aid for mobilization
  and survival}.
\newblock \bibinfo{journal}{\emph{Social Text}} \bibinfo{volume}{38},
  \bibinfo{number}{1} (\bibinfo{year}{2020}), \bibinfo{pages}{131--151}.
\newblock


\bibitem[Strengers et~al\mbox{.}(2021)]%
        {strengers2021can}
\bibfield{author}{\bibinfo{person}{Yolande Strengers}, \bibinfo{person}{Jathan
  Sadowski}, \bibinfo{person}{Zhuying Li}, \bibinfo{person}{Anna Shimshak},
  {and} \bibinfo{person}{Florian 'Floyd'~Mueller}.}
  \bibinfo{year}{2021}\natexlab{}.
\newblock \showarticletitle{What Can HCI Learn from Sexual Consent? A Feminist
  Process of Embodied Consent for Interactions with Emerging Technologies}. In
  \bibinfo{booktitle}{\emph{Proceedings of the 2021 CHI Conference on Human
  Factors in Computing Systems}} (Yokohama, Japan) \emph{(\bibinfo{series}{CHI
  '21})}. \bibinfo{publisher}{Association for Computing Machinery},
  \bibinfo{address}{New York, NY, USA}, Article \bibinfo{articleno}{405},
  \bibinfo{numpages}{13}~pages.
\newblock
\showISBNx{9781450380966}
\urldef\tempurl%
\url{https://doi.org/10.1145/3411764.3445107}
\showDOI{\tempurl}


\bibitem[Suresh et~al\mbox{.}(2022)]%
        {suresh22pml}
\bibfield{author}{\bibinfo{person}{Harini Suresh}, \bibinfo{person}{Rajiv
  Movva}, \bibinfo{person}{Amelia~Lee Dogan}, \bibinfo{person}{Rahul Bhargava},
  \bibinfo{person}{Isadora Cruxen}, \bibinfo{person}{Angeles~Martinez Cuba},
  \bibinfo{person}{Guilia Taurino}, \bibinfo{person}{Wonyoung So}, {and}
  \bibinfo{person}{Catherine D'Ignazio}.} \bibinfo{year}{2022}\natexlab{}.
\newblock \showarticletitle{Towards Intersectional Feminist and Participatory
  ML: A Case Study in Supporting Feminicide Counterdata Collection}. In
  \bibinfo{booktitle}{\emph{2022 ACM Conference on Fairness, Accountability,
  and Transparency}} (Seoul, Republic of Korea) \emph{(\bibinfo{series}{FAccT
  '22})}. \bibinfo{publisher}{Association for Computing Machinery},
  \bibinfo{address}{New York, NY, USA}, \bibinfo{pages}{667–678}.
\newblock
\showISBNx{9781450393522}
\urldef\tempurl%
\url{https://doi.org/10.1145/3531146.3533132}
\showDOI{\tempurl}


\bibitem[Tomasev et~al\mbox{.}(2021)]%
        {fairnessqueercomm}
\bibfield{author}{\bibinfo{person}{Nenad Tomasev}, \bibinfo{person}{Kevin~R.
  McKee}, \bibinfo{person}{Jackie Kay}, {and} \bibinfo{person}{Shakir
  Mohamed}.} \bibinfo{year}{2021}\natexlab{}.
\newblock \showarticletitle{Fairness for Unobserved Characteristics: Insights
  from Technological Impacts on Queer Communities}. In
  \bibinfo{booktitle}{\emph{Proceedings of the 2021 AAAI/ACM Conference on AI,
  Ethics, and Society}} (Virtual Event, USA) \emph{(\bibinfo{series}{AIES
  '21})}. \bibinfo{publisher}{Association for Computing Machinery},
  \bibinfo{address}{New York, NY, USA}, \bibinfo{pages}{254–265}.
\newblock
\showISBNx{9781450384735}
\urldef\tempurl%
\url{https://doi.org/10.1145/3461702.3462540}
\showDOI{\tempurl}


\bibitem[Wang et~al\mbox{.}(2023)]%
        {wang2023robustness}
\bibfield{author}{\bibinfo{person}{Jindong Wang}, \bibinfo{person}{Xixu HU},
  \bibinfo{person}{Wenxin Hou}, \bibinfo{person}{Hao Chen},
  \bibinfo{person}{Runkai Zheng}, \bibinfo{person}{Yidong Wang},
  \bibinfo{person}{Linyi Yang}, \bibinfo{person}{Wei Ye},
  \bibinfo{person}{Haojun Huang}, \bibinfo{person}{Xiubo Geng},
  \bibinfo{person}{Binxing Jiao}, \bibinfo{person}{Yue Zhang}, {and}
  \bibinfo{person}{Xing Xie}.} \bibinfo{year}{2023}\natexlab{}.
\newblock \bibinfo{title}{On the Robustness of Chat{GPT}: An Adversarial and
  Out-of-distribution Perspective}.
\newblock
\newblock
\urldef\tempurl%
\url{https://openreview.net/forum?id=uw6HSkgoM29}
\showURL{%
\tempurl}


\bibitem[Whittaker(2021)]%
        {whittaker2021steep}
\bibfield{author}{\bibinfo{person}{Meredith Whittaker}.}
  \bibinfo{year}{2021}\natexlab{}.
\newblock \showarticletitle{The Steep Cost of Capture}.
\newblock \bibinfo{journal}{\emph{Interactions}} \bibinfo{volume}{28},
  \bibinfo{number}{6} (\bibinfo{date}{nov} \bibinfo{year}{2021}),
  \bibinfo{pages}{50–55}.
\newblock
\showISSN{1072-5520}
\urldef\tempurl%
\url{https://doi.org/10.1145/3488666}
\showDOI{\tempurl}


\end{thebibliography}

\end{document}